\documentclass[journal]{IEEEtran}
\usepackage{cite}

\ifCLASSINFOpdf
  \usepackage[pdftex]{graphicx}
\else
\fi
\usepackage{booktabs}
\usepackage{multirow}
\usepackage{makecell}
\usepackage{amsmath}

\usepackage{stfloats}
\begin{document}
%
\title{Ultra-low Power AMOLED Displays for Smart Wearable Applications: Theory and Practice}
%
%
%

\author{Bojia Lyu
\thanks{B. Lyu is with the School of Electronic Information and  Electrical Engineering, Shanghai Jiao Tong University, Shanghai, 200240 China, and also with the Research and Development Center, Shanghai Tianma Microelectronics Co., Ltd, Shanghai, 201201 China.}
\thanks{e-mail: bojia.lyu@sjtu.edu.cn or bojia\_lv@tianma.cn}}


\maketitle

\begin{abstract}
With the continuous advancement and maturity of AMOLED (Active-Matrix Organic Light Emitting Diode) technology, smart wearable products such as watches and bracelets are increasingly incorporating related technologies as display screen implementation solutions. Using standby time is the most critical product performance measurement indicator at the moment, according to the power supply system design of smart wearable products and customer usage habits. AMOLED displays, as one of the major power-consuming components in smart wearable products, are also subject to extremely stringent power consumption requirements. This paper divides an AMOLED display into five parts: the power chip, the driver chip, the array substrate, the light-emitting structure, and the light-transmitting structure. In this paper, we propose targeted power-saving solutions for each component based on their respective operating principles, subject areas, and the most recent advances in related fields, and we provide the best overall solution by combining the interactions between each component and even the entire system. The relevant solutions have been validated in practice, and there is clear verification data to demonstrate their feasibility.
\end{abstract}


\IEEEpeerreviewmaketitle

\section{Introduction}
%
%
%
%
\IEEEPARstart{A}{fter} 
more than ten years of continuous development and practice, AMOLED display technology began to gradually replace traditional TFT-LCD (Thin Film Transistor Liquid Crystal Display) technology in various applications, further propelling the display industry's development \cite{1,2,3}. AMOLED products are being mass produced, particularly in the field of consumer electronics applications, such as television, cell phones, and smart wearables.

Each application branch has its own set of characteristics and requirements. People enjoy the excellent display effect brought by AMOLED technology at the same time, engineering and technical personnel also need to face and overcome the relevant areas in the matching AMOLED when encountering adaptation problems, and even need to combine the specific requirements of each field of AMOLED product design for targeted adjustments. Due to people's usage habits and product design architecture constraints, the power consumption of AMOLED products is the ultimate demand in the smart wearable field.

Smart wearable products are currently divided into two categories: smart watches and smart bracelets, which are worn on people's wrists instead of traditional watches. When compared to traditional watches, smart wearable products provide additional features such as mobile communication and sports health \cite{4,5,6}. People associate smart wearable products with traditional watch usage habits, such as light weight, small size, and long standby time.

Because light and compact products are small in size, the battery's design space is limited. In the context of limited battery capacity, achieving ultra-long standby is undoubtedly a challenge for the power consumption of smart wearable products. And, because AMOLED displays consume the most power in smart wearable products, power consumption has become the first evaluation index for AMOLED products in the field of smart wear.
\begin{figure}[!b]
\centering
\includegraphics[scale=0.7]{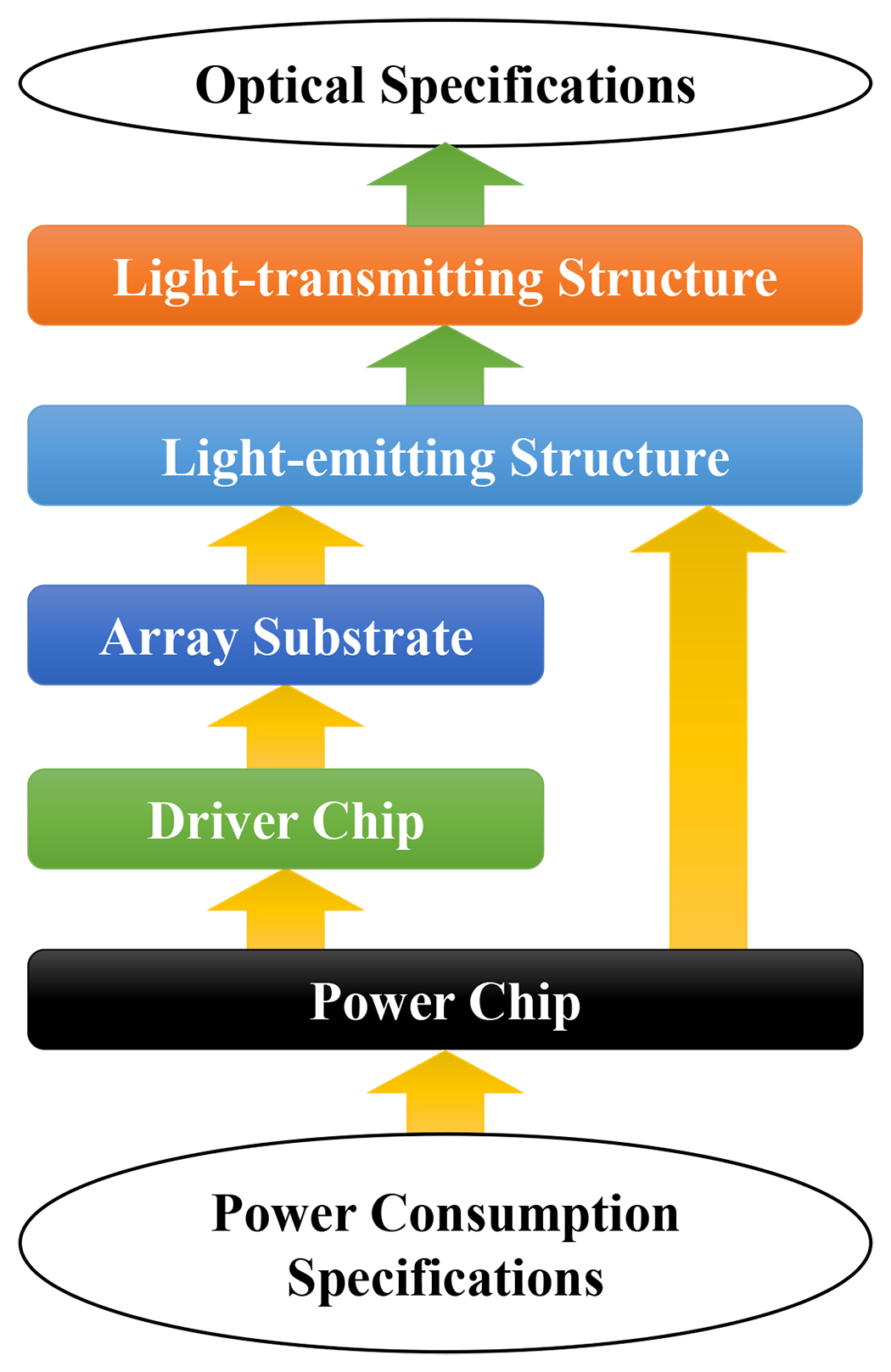}
\caption{AMOLED display system components for smart wearable application}
\label{fig:1}
\end{figure} 

One of the usage characteristics of smart wearable products is that they are worn for an extended period of time, but consumers only operate them for a short period of time. The AMOLED display has two application scenarios: the short time operation state, which requires the screen to be in normal display mode, and the long time standby state, where the screen only displays limited content or even no display. The research of ultra-low power AMOLED displays for smart wearable applications is primarily based on the two application scenarios mentioned above, and systematic development and design are carried out.

AMOLED products are best understood as an electro-optical conversion system (Figure \ref{fig:1}), which consists of five major components: a power chip, a driver chip, an array substrate, a light-emitting structure, and a light-transmitting structure. Power consumption is an indicator that evaluates the level of energy consumption at the power input and is closely related to each component of the conversion system. Under the premise of meeting the optical and performance requirements of the final output, the main means of studying ultra-low power AMOLED is how to design, deploy, and optimize each component to minimize the power consumption level at the power input.

\begin{table}[!b]
\caption{Optical specifications of AMOLED display for smart wearable application}\label{tab:1}
\centering
\begin{tabular}{cccccc}
\toprule
\multicolumn{2}{c}{ \multirow{2}{*}{Item} }& \multirow{2}{*}{Unit}& \multicolumn{3}{c}{Optical Specifications}\\

{}&{}&{}&max.&typ.&min.\\
\midrule
\multicolumn{2}{c}{Luminance (Normal)}& $cd/m^2$& {}& 600& 550\\
\multicolumn{2}{c}{Luminance (HBM)}& $cd/m^2$& {}& 1,000& 900\\
\multicolumn{2}{c}{Contrast Ratio}& -& {}& {}& 100,000\\
\multirow{8}{*}{\makecell[c]{Color\\Coordinate\\(CIE1931)}}& $x_{W}$& -& 0.309& 0.299& 0.289\\
{}& $y_{W}$& -& 0.325& 0.315& 0.305\\
{}& $x_{R}$& -& 0.707& 0.687& 0.667\\
{}& $y_{R}$& -& 0.335& 0.315& 0.295\\
{}& $x_{G}$& -& 0.297& 0.247& 0.197\\
{}& $y_{G}$& -& 0.766& 0.716& 0.666\\
{}& $x_{B}$& -& 0.171& 0.136& 0.101\\
{}& $y_{B}$& -& 0.085& 0.050& 0.015\\
\multicolumn{2}{c}{Uniformity}& \%& {}& {}& 85\\
\multicolumn{2}{c}{Color Uniformity}& -& 0.007& {}& {}\\
\multicolumn{2}{c}{NTSC}& \%& {}& 100& 97\\
\multicolumn{2}{c}{Crosstalk}& -& 1.1& {}& {}\\
\multicolumn{2}{c}{Color Temperature}& \emph{K}& 8,000& 7,500& 7,000\\
\multicolumn{2}{c}{Flicker (60Hz/G255)}& \emph{db}& -30& {}& {}\\
\multicolumn{2}{c}{Flicker (60Hz/G128)}& \emph{db}& -30& {}& {}\\
\multicolumn{2}{c}{Flicker (60Hz/G64)}& \emph{db}& -30& {}& {}\\
\multicolumn{2}{c}{Gamma}& -& 2.5& 2.2& 1.9\\
\multicolumn{2}{c}{Reflectance with Lens}& \%& 6& 5.5& {}\\
\multirow{4}{*}{\makecell[c]{Transmittance \\with Lens}}& 450nm& \%& {}& {}& 1\\
{}& 550nm& \%& {}& {}& 2.5\\
{}& 650nm& \%& {}& {}& 3\\
{}& 750nm& \%& {}& {}& 3\\
\multicolumn{2}{c}{Luminance decrease ratio}& \%& 40& 35& {}\\
\multirow{4}{*}{Color shift}& W& -& 4& 3& {}\\
{}& R& -& {}& {}& {}\\
{}& G& -& {}& {}& {}\\
{}& B& -& {}& {}& {}\\
\multicolumn{2}{c}{Response time}& \emph{ms}& 3& 2& {}\\
\multirow{4}{*}{\makecell[c]{Viewing \\Angle (85º) }}& L& -& {}& {}& 10\\
{}& R& -& {}& {}& 10\\
{}& T& -& {}& {}& 10\\
{}& B& -& {}& {}& 10\\
\multicolumn{2}{c}{Lifetime}& \emph{hr}& {}& {}& 300\\
\multicolumn{2}{c}{Image Retention}& \emph{s}& 5& {}& {}\\
\bottomrule
\end{tabular}
\end{table}

Although the power consumption specification appears to be an electrical indicator, it is actually the result of a number of factors at the product design and process levels. It is inextricably linked to the five AMOLED product components listed above and involves the intersection of multiple disciplines, making it a complex issue. In this paper, we will examine each of the five components listed above one by one, based on the path of signal transmission and transformation. We will explain the linkage and cooperation of upstream and downstream related components, in addition to introducing power consumption improvement solutions based on the characteristics of each component.

\subsection{Optical Specifications}
The wearable product display output conforms to the optical characteristics (Table \ref{tab:1}) and can be evaluated using a variety of optical specifications such as brightness, chromaticity, contrast, and so on. 

The final output result can be seen as the product's optical specifications. When designing, the above five components' material selection and program design are usually planned with the end in mind, but the power consumption specifications of the input side are frequently ignored.

When the relevant optical specifications have been met, power consumption calculations, evaluations, and improvements are performed to ensure that any optimization of power consumption does not result in changes to optical specifications or even exceeds customer minimum requirements.

On the contrary, if optimizing power consumption inevitably results in some optical specifications exceeding customers' minimum requirements, the necessity and reasonableness of the relevant specifications in the application of wearable products can be further explored, and if the corresponding requirements can be relaxed, it is equivalent to optimizing the product's power consumption in disguise.


\subsection{Power Consumption Specifications}

The input content of the wearable product display corresponds to the electrical characteristics, which include analog and digital power supplies, as well as display and control signals. The total power consumption of the analog and digital power supplies is the power consumption of the wearable product. Display and control signals involved in power consumption loss are generally subsumed into the input side is also the overall motherboard's output power consumption.

\begin{figure*}[htbp]
\centering
\includegraphics[width=5in]{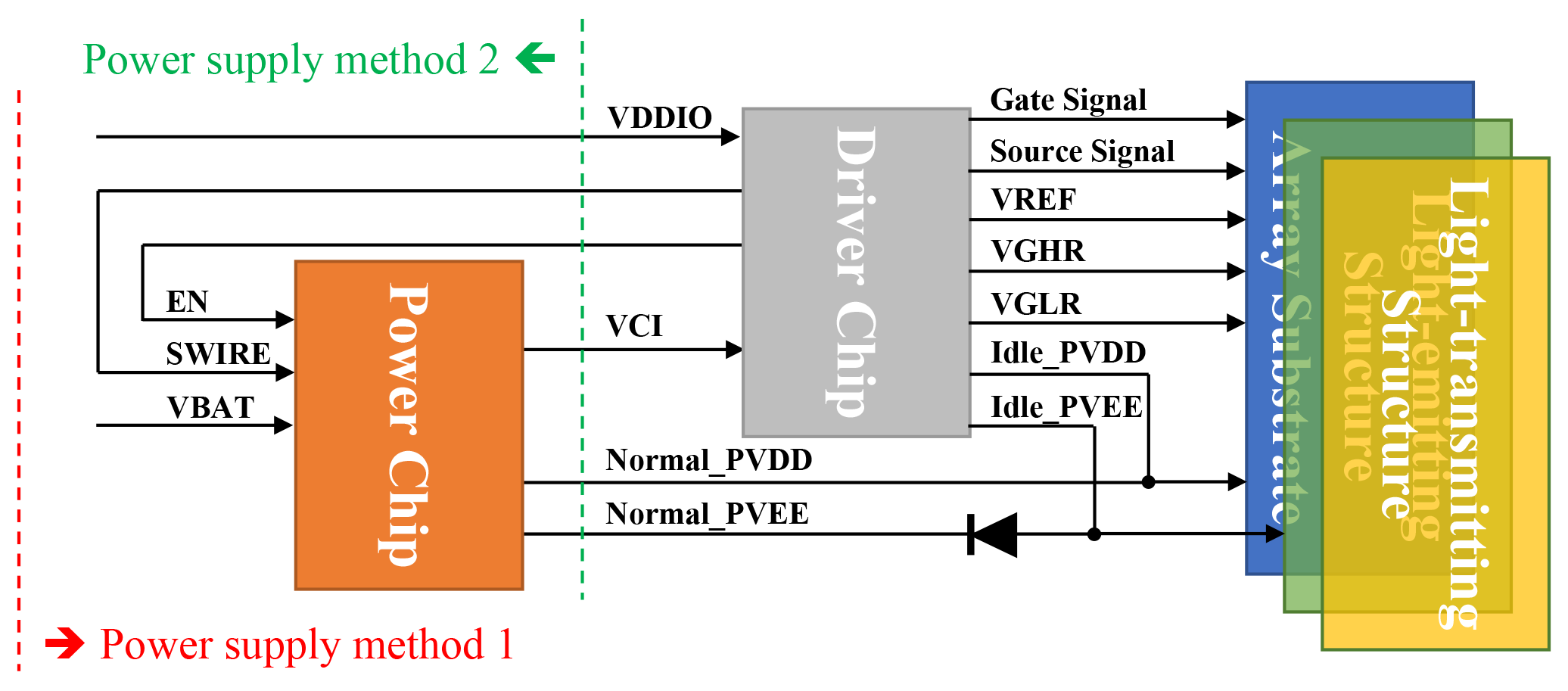}
\caption{AMOLED display system components for smart wearable application}
\label{fig:2}
\end{figure*}

The calculation of display power consumption differs depending on the location of the power supply chip, but the logic remains the same. The power consumption can be calculated in two methods, depending on the connection between the power supply chip and the driver chip (Figure \ref{fig:2}).
    
Power supply method 1: The power supply chip is located on the display module

\begin{equation}\label{eq1}
    P=V_{DDIO}\times I_{DDIO}+V_{BT}\times I_{BT}
\end{equation}

Power supply method 2: The power supply chip is located on the customer's motherboard

\begin{equation}\label{eq2}
\begin{aligned}
P=&V_{DDIO}\times I_{DDIO}+V_{CI}\times I_{CI}+
\\
&(V_{PVDD}-V_{PVEE}\ )\times I_{PVEE}
\end{aligned}
\end{equation}

Because wearable devices are rarely used for video playback in their applications, display data is typically only passed on when the display screen's content needs to be changed. The content of the screen displayed for an extended period of time will be stored in the driver chip's memory, ensuring that the display can work at its own required refresh rate. This mode of operation saves power on both the display and the entire wearer, and it has become the industry standard for use. As a result, this data transfer mode is used in the power consumption calculation and specification determination.

\subsection{Evaluation Criteria}
AMOLED displays have four different operating modes, each with different operating states and power consumption requirements (Table 2), depending on the application scenarios of smart wearable products.

\begin{table}[!b]
\caption{Power consumption specifications (CPK \textgreater 1.33) of 1.19inch AMOLED display (power supply method 1, VDDIO = 1.8V, VBAT = 3.7V)}\label{tab:2}
\centering
\begin{tabular}{ccccc}
\toprule
\multirow{2}{*}{Condition}& \multirow{2}{*}{Mode}& \multicolumn{3}{c}{\makecell[c]{Power Consumption \\Specifications (mW)}}\\

{}&{}&max.&typ.&min.\\
\midrule
100\% Pixel on, 450nits, 60Hz& Normal& {}& 150& 160\\
100\% Pixel on, 450nits, 45Hz& Normal& {}& 145& 156\\
100\% Pixel on, 450nits, 30Hz& Normal& {}& 142& 155\\
All pixel off, 60Hz& Normal& {}& 13& 15\\
All pixel off, $V_{CI}$ on, $V_{DDIO}$ on& Standby& {}& 1.0& 1.2\\
All pixel off, $V_{CI}$ off, $V_{DDIO}$ on& Standby& {}& {}& 0.0016\\
10\% Pixel on, 50nits, 15Hz& Idle& {}& 7.6& 8.4\\
10\% Pixel on, 100nits, 15Hz& Idle& {}& 10.2& 11.2\\
100\% Pixel on, 1000nits, 60Hz& Boost& {}& 350& 410\\
\bottomrule
\end{tabular}
\end{table}

\subsubsection{Normal Mode}
Normal display mode refers to the display mode of the wearable product when it is used normally indoors. It must meet all of the optical characteristics and have all of the display functions, which are only visible during operation, and the daily use time is relatively short.
\subsubsection{Boost Mode}
Highlighting mode, which is used when the wearable product is used outside, or sweeping code function. The brightness is increased based on the Normal mode function setting, and the time spent in daily use is very short.
\subsubsection{Idle Mode}
Idle display mode, which is used when the wearable product is not in use. The optical properties are reduced, and some unnecessary display functions are disabled. Only limited and simple information is displayed, and the time spent in daily use is increased.
\subsubsection{Standby Mode}
Standby mode is used when the wearable product is not in use. Based on Idle mode, it further disables most functions, including the display function, and only retains the power to wait for waking up, which takes a very long time in daily use.

\section{Power Chip}
The power supply chip is a power management integrated circuit (PMIC, Power Management Integrated Circuit) with three-channel power output and voltage programmable function, and it is a critical component for powering the display module \cite{7, 8, 9}. Figure \ref{fig:2} demonstrates the connection of the power supply chip as well as the input and output signals in each component of AMOLED display products.

The power supply chip is powered by the battery assembly of the entire wearable product (VBAT) and, depending on the customer's needs, can be designed on the display module or on the customer's motherboard. When evaluating the power consumption of the display module alone, the difference in design location can lead to different results, but there is no difference in overall power consumption from the standpoint of the entire machine.

As shown in Figure \ref{fig:2}, the power supply chip primarily powers the driver chip (VCI) and the light-emitting structure (PVDD/PVEE). The power supply chip, as the core component of AMOLED display power supply, has a significant share in the power consumption specifications of different working modes of smart wearable products.

\begin{figure}[!t]
\centering
\includegraphics{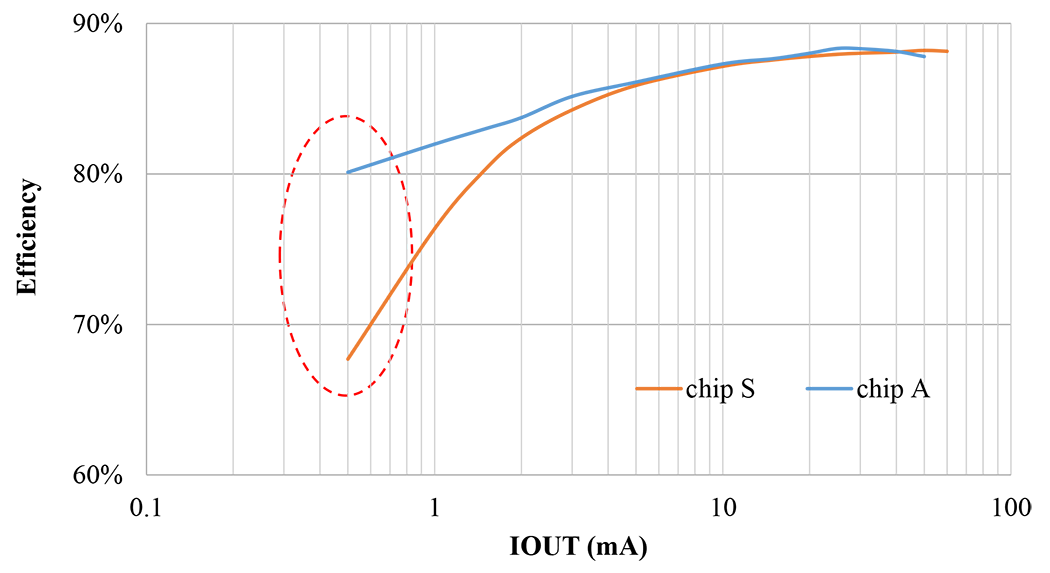}
\caption{Efficiency curves of power chip S and chip A at the same voltage (PVDD/PVEE=3.3V/-3.3V) setting.}
\label{fig:3}
\end{figure}

Different manufacturers' power supply chips are affected by their respective designs and processes, and there are differences in power conversion efficiency under the same operating mode and settings (Figure \ref{fig:3}). This is primarily reflected in the difference in efficiency under different load conditions, as well as the load difference corresponding to the optimal efficiency point. The best matching conditions must be chosen based on the specific design scheme and application conditions of the AMOLED display.

\begin{figure}[!t]
\centering
\includegraphics{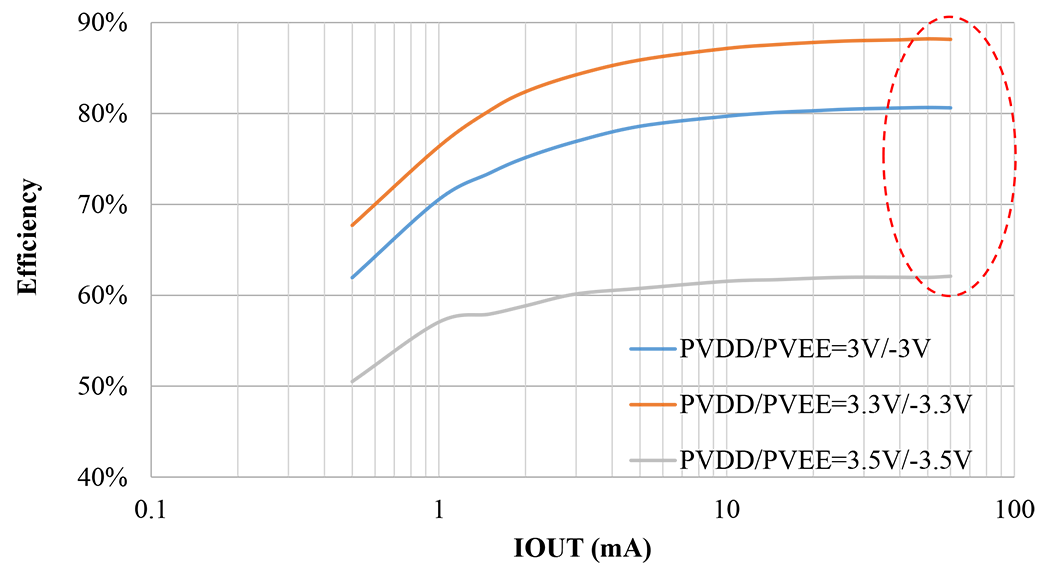}
\caption{Efficiency curves of power chip S at different voltage settings.}
\label{fig:4}
\end{figure} 

The efficiency curves of power supply chips at various voltage output settings differ, and the differences are significant (Figure \ref{fig:4}). The chip will be designed with demand for a specific voltage output setting in mind for efficiency optimization; deviation from the optimal settings will result in a decrease in efficiency. The selection of project opening evaluation chips must be combined with specific needs for targeted selection.

Customizing the design of power supply chips to meet project requirements for voltage output settings and common load ranges is an effective way to improve power supply chip conversion efficiency. However, customization of power supply chips requires a significant investment and a lengthy lead time, and in the current mature industrial product development process, power supply chips already on the market are frequently chosen.

As a result, the direction of this paper to focus on the next study is how to choose a power supply chip in many similar products, and how to use a good power supply chip. Power conversion efficiency is a key indicator of the power chip that influences the display's power consumption. If the conversion efficiency of an existing mature power supply chip cannot be improved, effective control of power loss is another way to improve conversion efficiency in disguise.

\subsection{Inductor Selection}
The inductor is a necessary component of the peripheral circuit of the current mainstream power supply chip and has a direct impact on the power supply chip's normal operation. When choosing a device, most people focus on the inductance value rather than the inductor's heat rating current, and DC resistance. However, the latter is the most important factor in determining the inductor's power consumption during operation.

\begin{figure}[!b]
\centering
\includegraphics{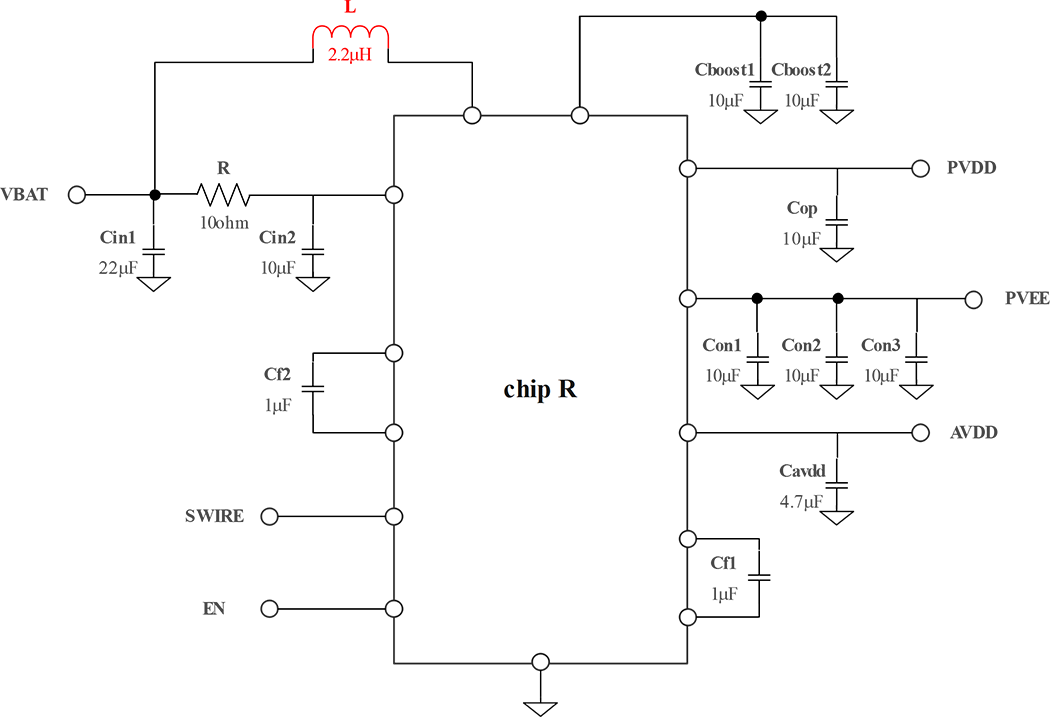}
\caption{Typical application circuit of power chip R, include capacitors, resistors and inductors.}
\label{fig:5}
\end{figure}

When the power supply chip is operational, peripheral devices must be connected to at least one inductor design (Figure \ref{fig:5}). Because the current flowing through the inductor will cause power loss due to its own characteristics \cite{10,11}, the maximum power loss calculation formula is:

\begin{equation}\label{eq3}
P_{DCR}= I_{rms}^2\times DCR
\end{equation}

where, $I_{rms}$ is the heat rating current of the inductor; $DCR$ is the DC resistance of the inductor.
When choosing an inductor, choose the model with the lowest power loss by matching $I_{rms}$ and $DCR$, while meeting the best matching inductor value provided by the power supply chip. The higher the loading in Normal and Boost mode, the greater the power loss caused by the inductor.

\begin{figure}[!b]
\centering
\includegraphics{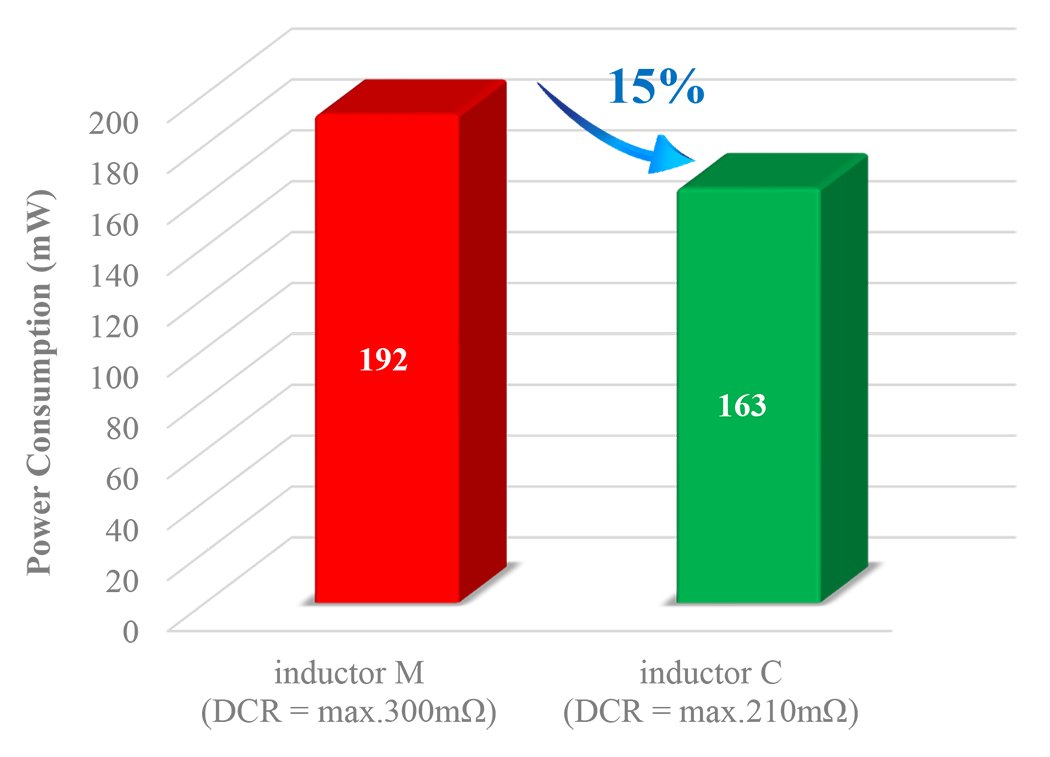}
\caption{The effect of inductors with different DCR specifications on the power consumption of one 1.19inch AMOLED display, under the condition of 100\% pixel on, 450nits and 60Hz.}
\label{fig:6}
\end{figure} 

On a 1.19" AMOLED display, we compared the power consumption of inductors with different DC resistances (Figure \ref{fig:6}). As a result, by replacing inductor devices with a smaller DC resistance, the power supply chip's conversion efficiency can be improved, and thus power consumption can be reduced, while maintaining the same design and settings.
\subsection{Quiescent Current}
The internal current consumed by the chip itself in the no-load state is referred to as quiescent current.

The power supply chip's quiescent current is very small, and the current size is closely related to the design of the integrated circuit and the semiconductor process \cite{12,13,14}. The quiescent current of the listed chips is currently fixed and cannot be increased by external means.

Choosing a power supply chip with a low quiescent current is an effective design method for reducing power loss. Idle and standby modes, in particular, such as low load mode, can significantly reduce power consumption.

At the same time, consider whether the chip supports work mode and standby mode switching; different modes of quiescent current often have significant differences that can be targeted to improve.

\subsection{Conversion Mode}

\begin{figure}[!b]
\centering
\includegraphics{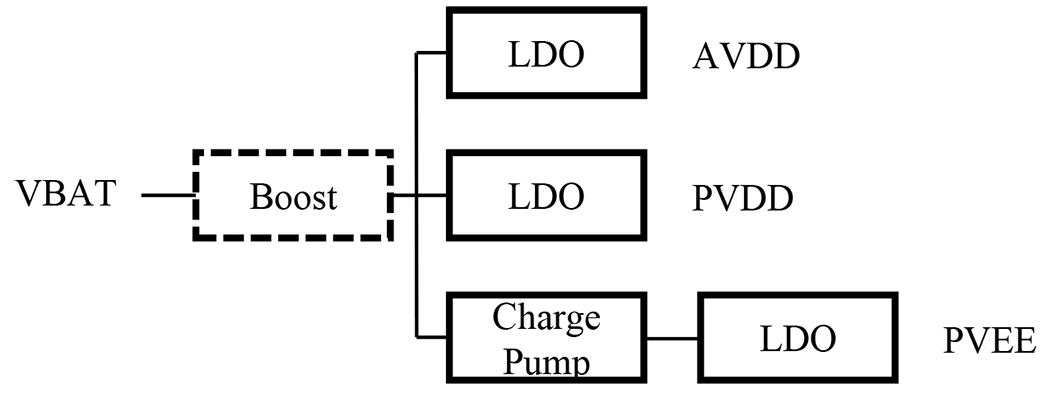}
\caption{Power chip internal voltage conversion module, include boost, charge bump and LDO.}
\label{fig:7}
\end{figure}

The power chip's primary function is to convert voltage input and output through various internal voltage conversion circuits to achieve the desired voltage output \cite{15,16}. This may necessitate the use of multiple internal conversion circuits, and each voltage must pass through two or three conversion circuits in order to produce the desired result (Figure \ref{fig:7}). Each voltage conversion circuit results in a power consumption loss, and the more conversion circuits that are passed, the greater the loss.

The power conversion efficiency can be improved by making some adjustments to the input and output in accordance with the power conversion principle. For example, eliminate or bypass the boost circuit if you can ensure that the input voltage is always greater than the output voltage. This can significantly improve overall conversion efficiency, particularly in normal and boost modes, such as high load mode.

\begin{figure}[!t]
\centering
\includegraphics[scale=0.9]{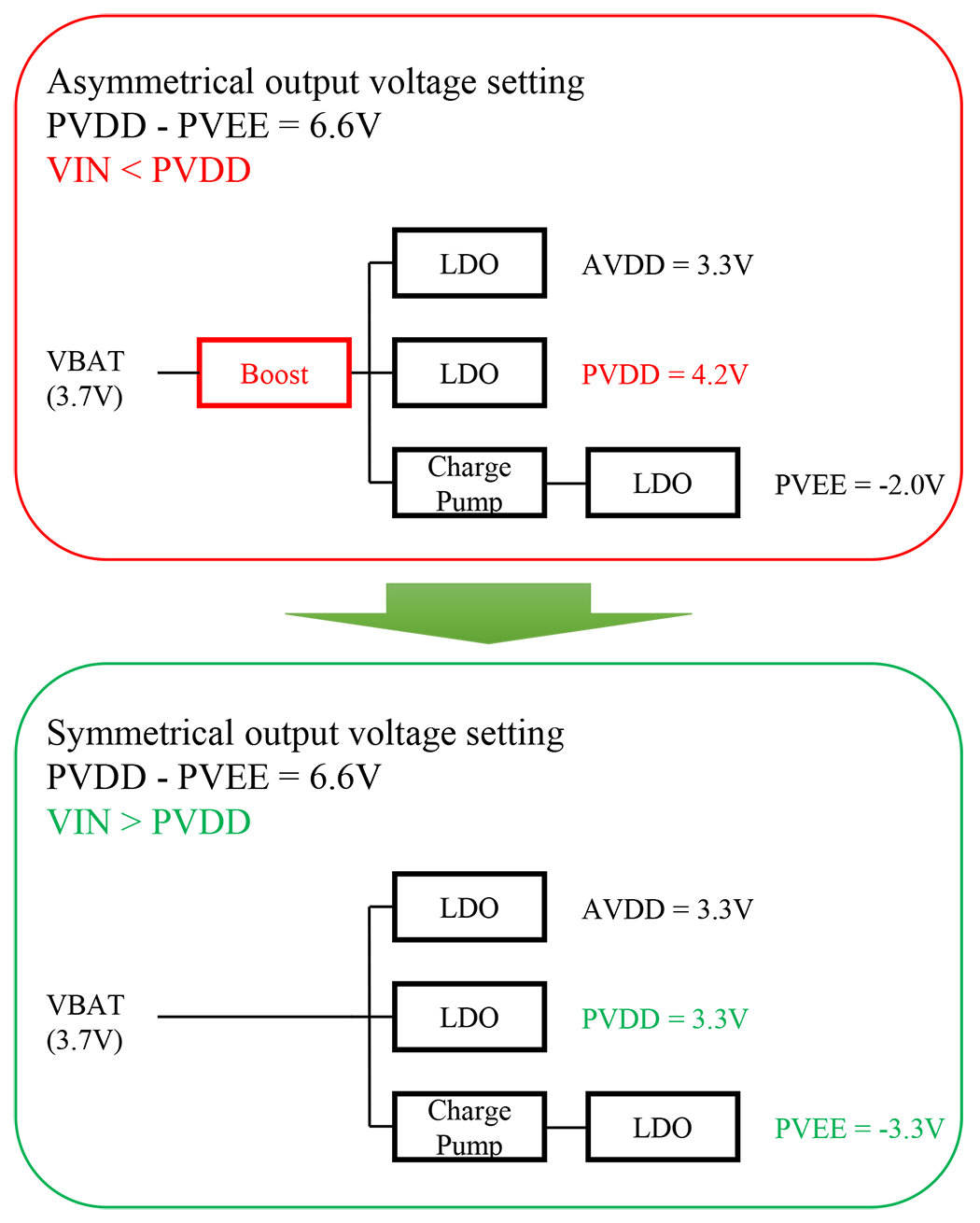}
\caption{Comparison of symmetrical and asymmetrical output voltage setting schemes for power chips, maintaining the voltage difference between PVDD and PVEE constant.}
\label{fig:8}
\end{figure}

\begin{figure}[!t]
\centering
\includegraphics[scale=0.9]{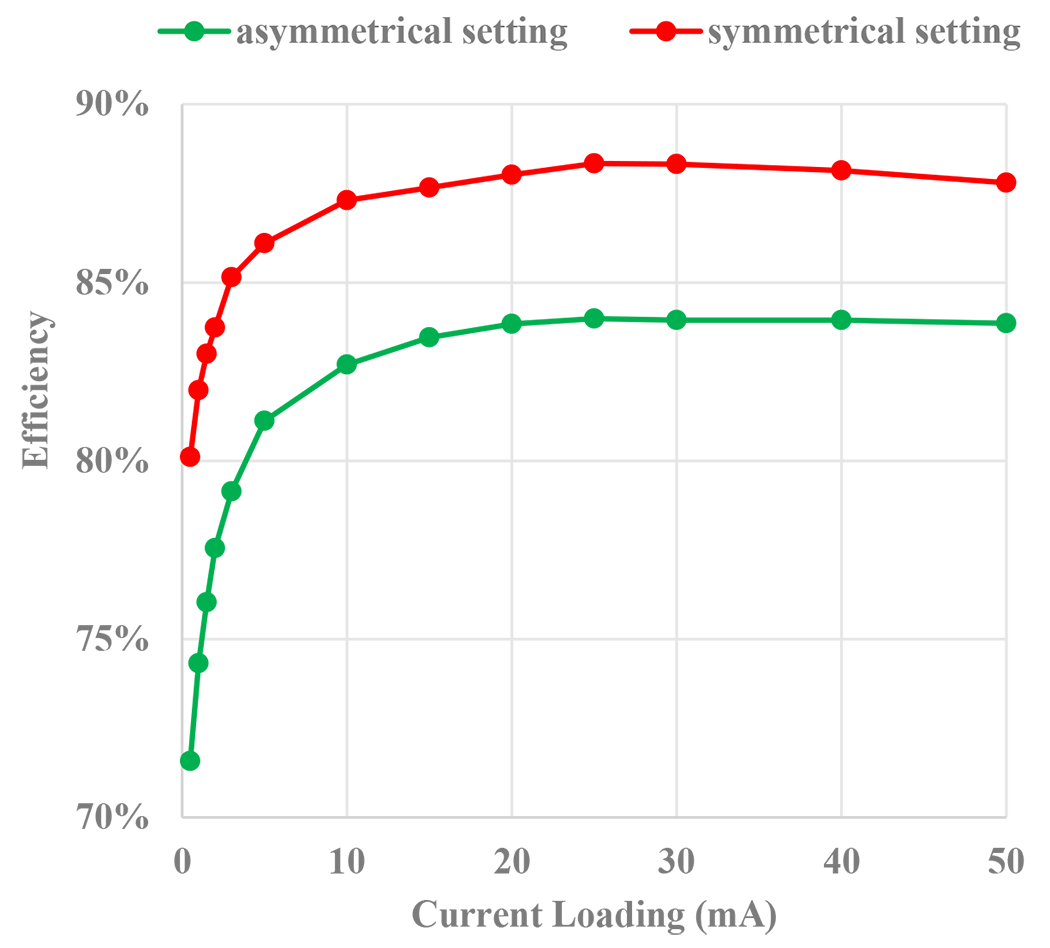}
\caption{Efficiency curves at symmetrical and asymmetrical output voltage settings for the power chip A, maintaining the voltage difference between PVDD and PVEE constant.}
\label{fig:9}
\end{figure}

By changing the asymmetric voltage output to a symmetric voltage output while keeping the PVDD/PVEE across the voltage constant, the power chip bypasses the boost voltage conversion circuit and increases power supply conversion efficiency (Figure \ref{fig:8}). According to the measured data (Figure \ref{fig:9}), the conversion efficiency of heavy load power supply is improved by nearly 5\% and light load power supply is improved by nearly 10\%.

\section{Driver Chip}

The driver chip is a display driver integrated circuit (DDIC) that combines data processing, digital-to-analog conversion, and other functions. It primarily drives and controls the array substrate's normal operation.

The internal system of the driver chip is divided into two parts: the digital circuit and the analog circuit, with independent input power supplies (Figure \ref{fig:2}). The power chip provides the analog power supply (VCI), while the customer's motherboard provides the digital power supply (VDDIO).

Because the driver chip is not in charge of powering the light-emitting structure in normal or boost mode, it does not account for a significant portion of power consumption in these two modes. However, in idle and standby mode, its own power consumption takes precedence.

\subsection{Analog Circuit Section}

\begin{figure*}[!b]
\centering
\includegraphics[scale=0.9]{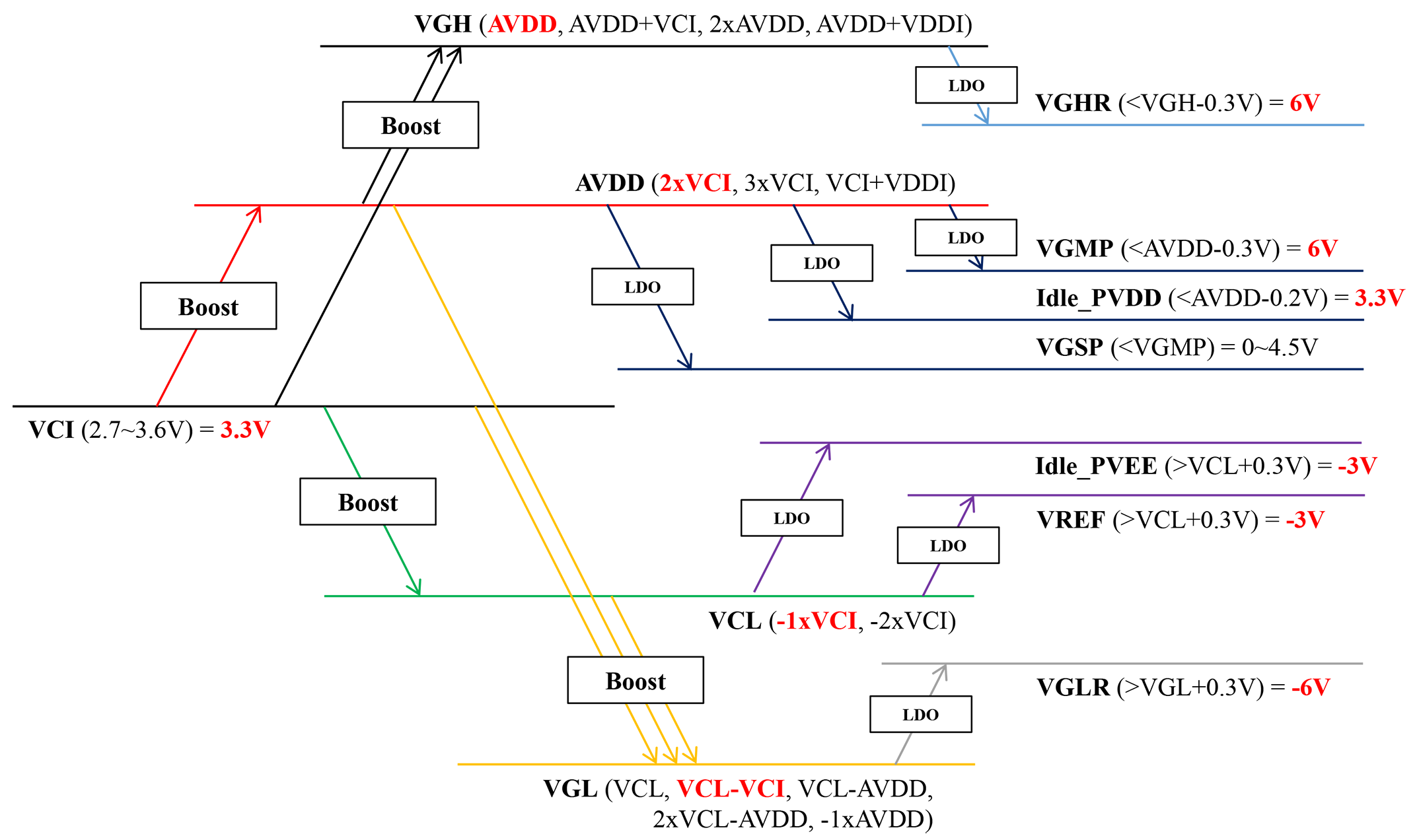}
\caption{Internal voltage generation diagram of driver chip.}
\label{fig:10}
\end{figure*}

The driver chip's function and design are far more complex than those of the power chip, where the analog part is primarily responsible for generating the working power required by the multiplexed array substrate. This part of the working principle is similar to that of the power supply chip, but the design is more complex and has more configuration options (Figure \ref{fig:10}). However, the power consumption improvement program and the power supply chip are the same in principle, with the main means being optimization of the boost mode configuration and voltage conversion mode.

The power conversion circuit within the driver chip is typically designed with different setting options based on demand, with the conversion efficiency of each option varying. In practice, based on demand, choosing the conversion efficiency of the higher setting options can effectively improve the level of power consumption. Furthermore, the less conversion unit passed in the conversion circuit selection and power supply chip, the less power loss.

\begin{table}[!t]
\caption{Internal voltage configuration of driver chips R in different boost modes}\label{tab:3}
\centering
\begin{tabular}{cccc}
\toprule
Voltage Setting& Case A& Case B& Case C\\
\midrule
VCI& 3.3V& 3.3V& 3.3V\\
\makecell[c]{AVDD \\ {\scriptsize $(2 \times VCI)$}}& 6.5V& 6.5V& 6.5V\\
\makecell[c]{VGMP \\ {\scriptsize $(\textless AVDD-0.3V)$}}& 5.4V& 5.6V& 5.4V\\
VGH& \makecell[c]{9.8V \\ {\scriptsize $(AVDD+VCI)$}}& \makecell[c]{6.5V \\ {\scriptsize $(AVDD)$}}& \makecell[c]{6.5V \\ {\scriptsize $(AVDD)$}}\\
\makecell[c]{VGHR \\ {\scriptsize $(\textless VGH-0.3V)$}}& 7V& 6V& 6V\\
VCL& \makecell[c]{-6.6V \\ {\scriptsize $(-2 \times VCI)$}}& \makecell[c]{-6.6V \\ {\scriptsize $(-2 \times VCI)$}}& \makecell[c]{-3.3V \\ {\scriptsize $(-1 \times VCI)$}}\\
\makecell[c]{VREF \\ {\scriptsize $(\textgreater VCL+0.3V)$}}& -3.5V& -3.5V& -2.5V\\
\makecell[c]{VGL \\ {\scriptsize $(VCL-VCI)$}}& -9.9V& -9.9V& -6.6V\\
\makecell[c]{VGLR \\ {\scriptsize $(\textgreater VGL+0.3V)$}}& -6V& -6V& -5V\\
\bottomrule
\end{tabular}
\end{table}

\subsubsection{Boost Mode Configuration}

Inside the driver chip, there are multiple boost voltage conversion circuits for internal boosting action, and each boost circuit has several different boost levels to choose from, making it easy to adapt to the needs of different types of array substrates. In the context of meeting the specific requirements of the array substrate voltage settings, selecting a lower boost level can help to reduce the driver chip's power consumption.

The premise of the optimal boost mode configuration is to not affect the array substrate's input voltage requirements, which affects the display effect under normal operation. If the voltage combination required by the array substrate cannot be further optimized in the current drive chip optional boost mode, the display effect should not be sacrificed in exchange for further power consumption reduction.

\begin{figure}[!b]
\centering
\includegraphics{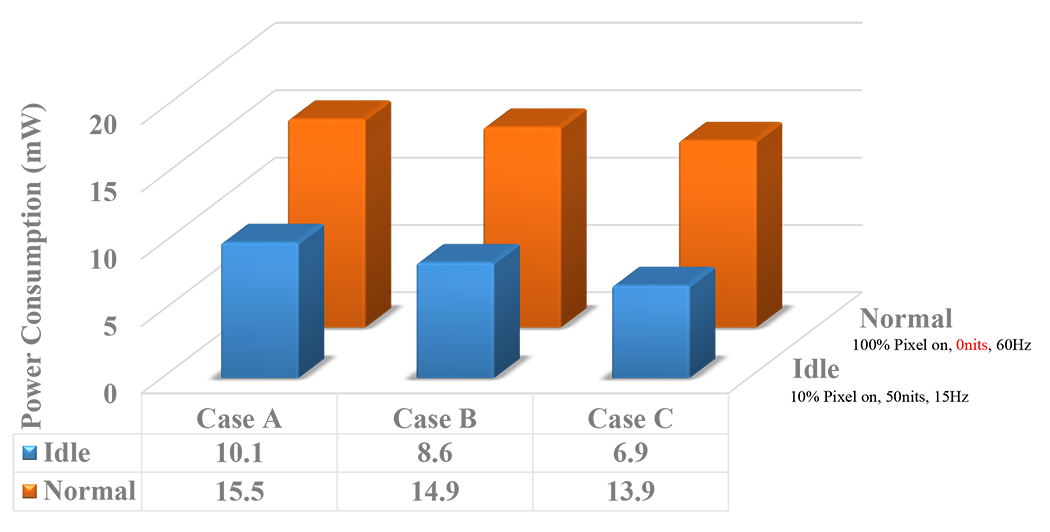}
\caption{The effect of driver chip R with different internal voltage boost modes on the power consumption of one 1.19inch AMOLED display, under the condition of Idle and Normal modes.}
\label{fig:11}
\end{figure}

By optimizing the boost mode configuration of the driver chip's voltage conversion circuit on the same AMOLED display (Table \ref{tab:3}), the impact on the power consumption level of the driver chip itself can be observed. When the power consumption test data of the two configurations before and after are compared (Figure \ref{fig:11}), it is discovered that the maximum power consumption reduction of the AMOLED display in normal mode (non-illuminated) is nearly 10\%, and the maximum power consumption reduction in idle mode is nearly 30\%.

\subsubsection{Voltage Conversion Mode}

The voltage of the driver chip, like the voltage of the power chip, will typically pass through multiple voltage conversion circuits from input to output. The greater the number of voltage conversion circuits passed, the greater the accumulated power loss. Some voltage conversion circuits can be bypassed to reduce power consumption when the input voltage is greater than or equal to the output voltage.

\begin{figure}[!t]
\centering
\includegraphics[scale=0.9]{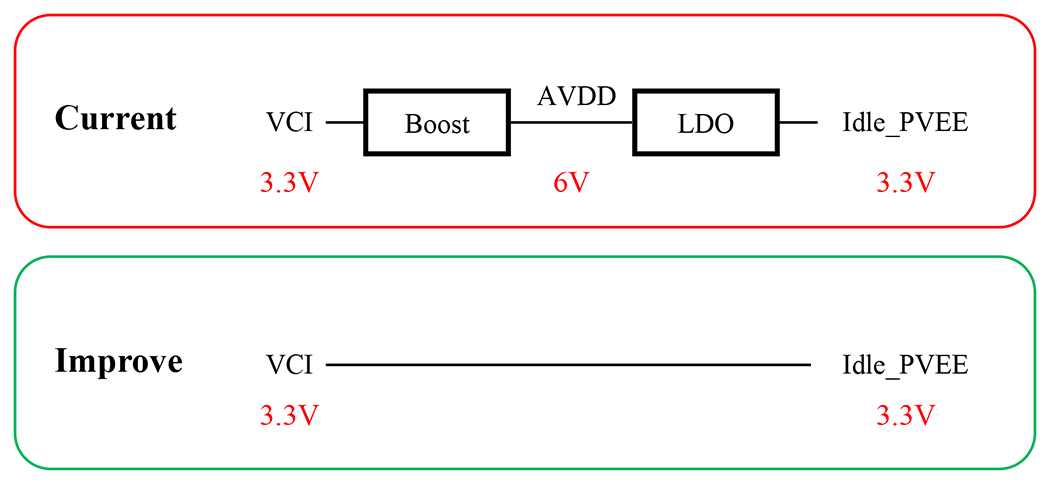}
\caption{Driver chip input and output voltage conversion streamlined solution}
\label{fig:12}
\end{figure}

\begin{figure}[!t]
\centering
\includegraphics{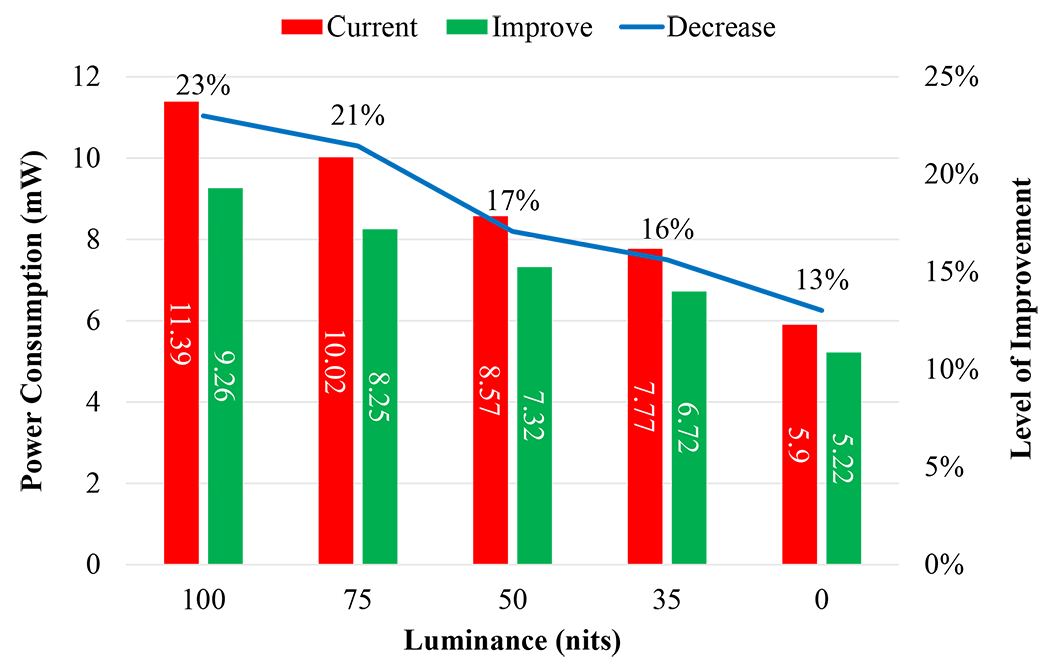}
\caption{The effect of driver chip N before and after improvement of input/output voltage conversion mode on power consumption of one 1.43inch AMOLED display, under the condition of 10\% pixel on and 15Hz with different luminance.}
\label{fig:13}
\end{figure}

The change in power consumption before and after changing the generation method (Figure \ref{fig:12}) of Idle\_PVEE output voltage on the same AMOLED display can be observed. By bypassing the boost and LDO voltage conversion circuits, the AMOLED display can improve its power consumption by close to 20\% under various loads (Figure \ref{fig:13}). However, the LDO voltage conversion circuit also ensures output voltage stability, so if you choose to bypass the LDO, you must thoroughly evaluate the impact of ripple increase on the display.

\subsection{Digital Circuit Section}
\begin{figure*}[!t]
\centering
\includegraphics{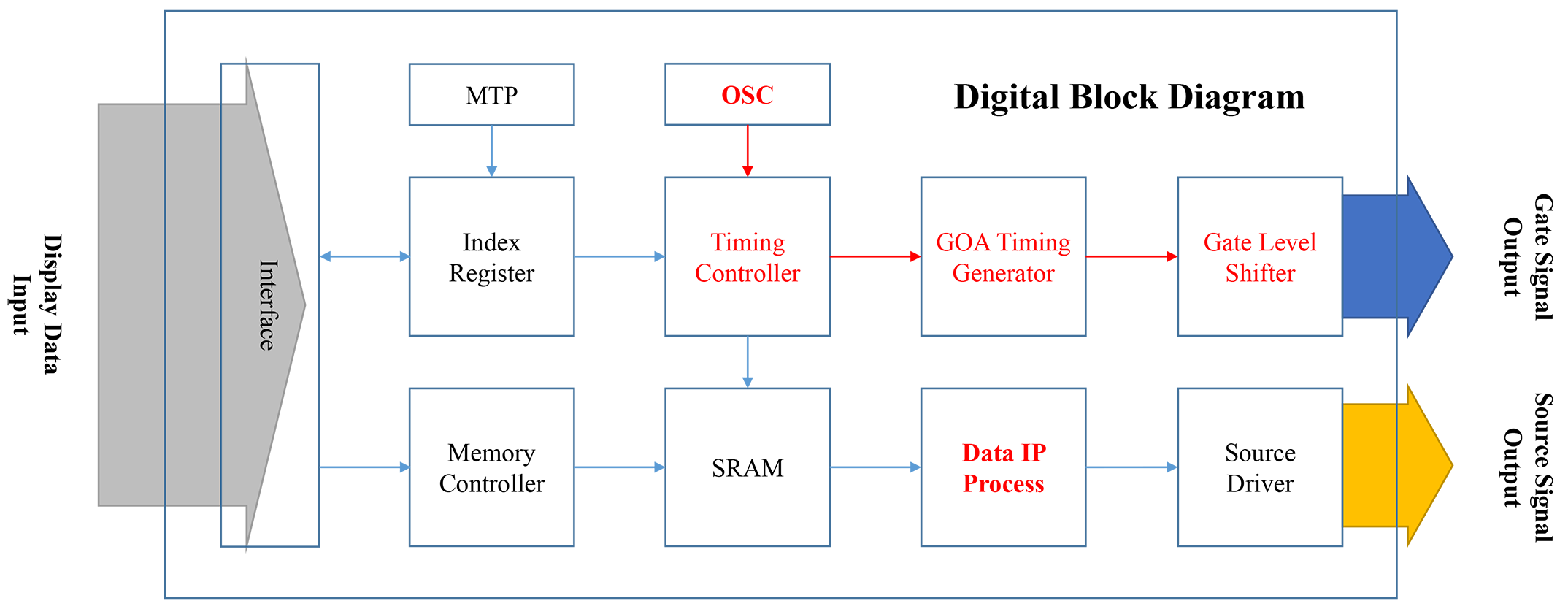}
\caption{Driver chip digital circuit function module.}
\label{fig:14}
\end{figure*}

The difference between a driver chip and a power chip in the design of analog circuit section is minimal, and even power optimization means that some can be shared. However, the driver chip contains the digital circuit section that includes data transmission, timing matching, image processing, digital-to-analog conversion, and other functional modules (Figure \ref{fig:14}).

Reducing the clock timing frequency and turning off non-essential function modules are the main ways to improve the power consumption of the driver chip's digital circuit under the premise of ensuring the basic working requirements of the display module and the display effect can be met.

The power consumption of the digital circuit accounts for a small portion of the total power consumption of the driver chip and even the display module, and the corresponding improvement means can only highlight the corresponding contribution in idle and standby mode.

\subsubsection{Clock Frequency Adjustment}
The clock frequency is the basis for driving the digital part of the chip to work. It involves and affects many different aspects, the most direct manifestation is the refresh frequency of the display screen. Standing in the perspective of the driver chip, the lower the clock frequency the lower power consumption, which is a key entry point for improvement of power consumption.

Clock frequencies are ubiquitous in the digital circuitry section of the driver chip. The clock frequency of the driver chip is related to everything from the oscillator's high frequency clock signal to the screen refresh's low frequency clock signal. It is a good idea to reduce the frequency of the driver chip to reduce power consumption, but the choice of optimization object determines the degree of improvement in power consumption.

The power consumption improvement will bring, whether it is to reduce the oscillation frequency of the chip or the refresh frequency of the screen. However, because the display components involved in both are different, the results of power consumption improvement differ. The oscillation frequency serves as the foundation for driving the chip clock frequency, but it is only used by the internal drive chip. The refresh frequency affects the entire display and involves many factors.

\begin{figure}[!b]
\centering
\includegraphics{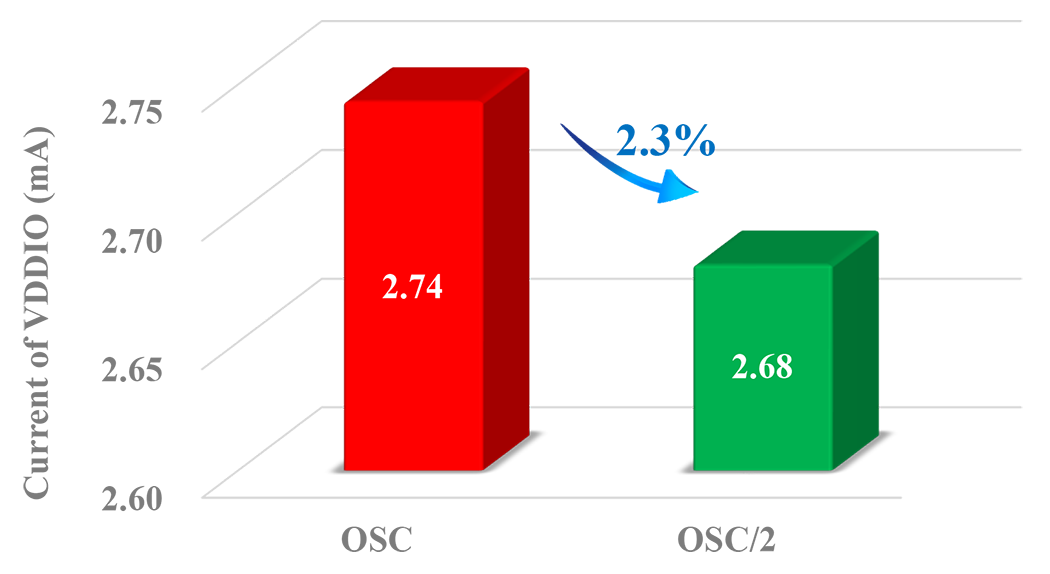}
\caption{The effect of driver chip C with different oscillation frequency setting on power consumption of one 1.43inch AMOLED display, under the condition of 100\% pixel on, 0nit and 60Hz.}
\label{fig:15}
\end{figure}

In the case of satisfying the driver chip's normal operation, appropriately lowering the oscillation frequency has no effect on the output results. The difference in power consumption before and after the improvement of the same AMOLED product is observed by halving the oscillation frequency of the driver chip (Figure \ref{fig:15}). The relevant improvement can result in a 2.3\% reduction in power consumption, all of which is due to an improvement in the power consumption of the digital circuit part of the driver chip itself.

\begin{figure}[!b]
\centering
\includegraphics{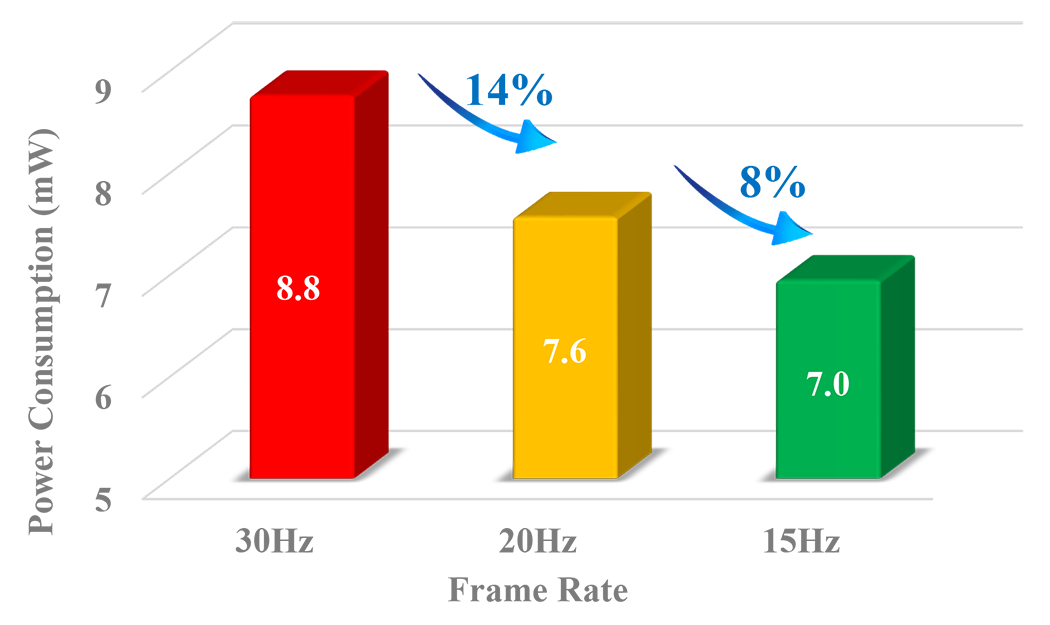}
\caption{The effect of driver chip R with different frame frequency setting on power consumption of one 1.19inch AMOLED display, under the condition of 10\% pixel on and 50nits.}
\label{fig:16}
\end{figure}

Screen refresh frequency is an important indicator of AMOLED displays and plays an important role in product display. It affects not only the digital circuit part of the driver chip, but also the analog circuit part and the array substrate. The change in power consumption of the same AMOLED product with different screen refresh frequencies (Figure \ref{fig:16}) reveals that the resulting load changes have a direct impact on the power consumption results.

\subsubsection{Functional Module Adjustment}

The driver chip's image processing is a critical component of its digital circuit section. The driver chip generally has a variety of image processing algorithms built in to facilitate customer selection based on the various functions and needs. Most image processing functions are usually enabled by default in order to achieve excellent display effects. However, when combined with the various operating modes of wearable devices, you can save power by turning off unnecessary image processing functions.

Only the Standby mode of the wearable product does not require the display of the screen and can turn off all digital circuits and even some analog circuits to achieve the ultimate power-saving purpose. Boost, Normal, and Idle modes, on the other hand, will require a screen display and can only be judged by the difference in display effect requirements in their respective modes to determine whether there is room for optimization.

\begin{figure}[!t]
\centering
\includegraphics{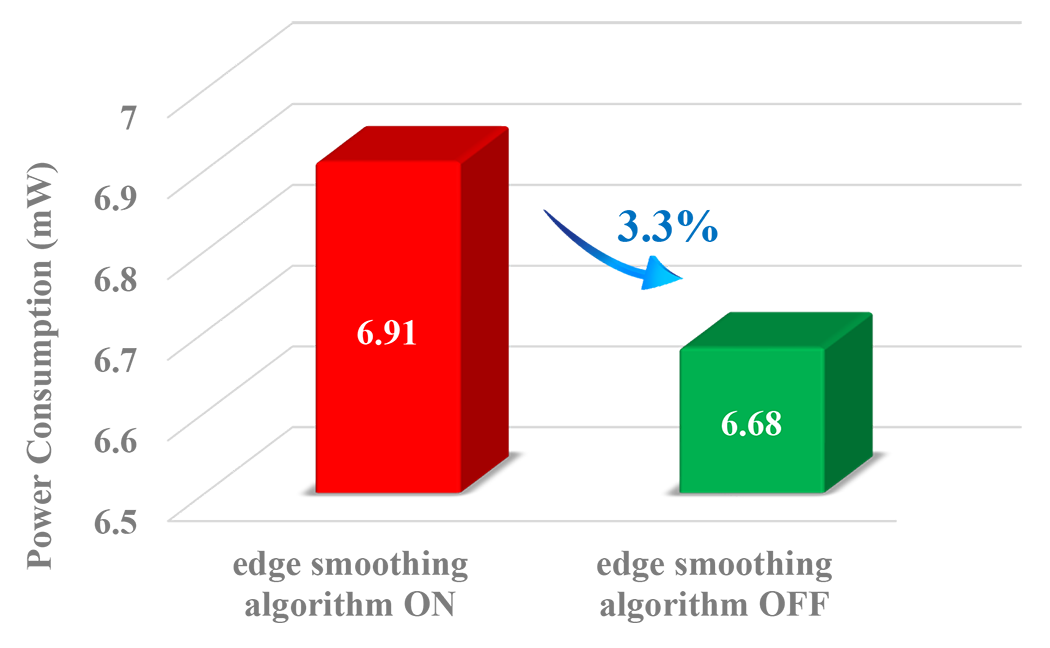}
\caption{The effect of driver chip R with and without edge smoothing algorithm on power consumption of one 1.19inch AMOLED display, under the condition of 10\% pixel on, 50nits and 15Hz.}
\label{fig:17}
\end{figure}

The display edge smoothing algorithm \cite{17}, for example, is an image processing function in the driver chip that eliminates jagged edges of shaped displays, but it is only available in full-screen mode. Because there is no need for full-screen display in Idle mode, turning off the edge algorithm has no effect on the usage effect. A 3.3\% improvement in power consumption can be obtained by comparing the power consumption difference between the same AMOLED product with the edge smoothing algorithm on and off in idle mode (Figure \ref{fig:17}).

\subsection{Linkage with Power Chip}

The driver and power chips do not operate independently. There is information interaction and timing linkage between them. However, if you use commercially available chips for display design, you will run into the issue of coordination linkage between different brands and models. The focus of this chapter is on how to achieve a reasonable configuration between the driver chip and the power chip so that AMOLED displays in different operating modes can achieve the best power consumption level.

\begin{figure}[!b]
\centering
\includegraphics{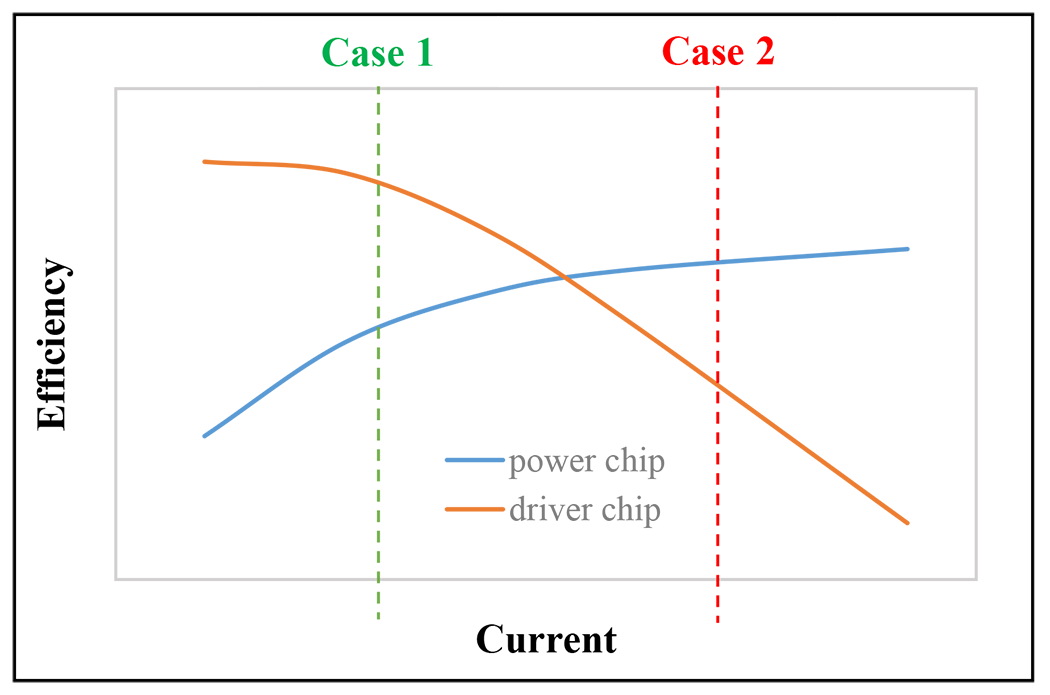}
\caption{Schematic curve of power conversion efficiency under the light loading of driver and power chip.}
\label{fig:18}
\end{figure} 

\subsubsection{Hybrid Power Supply Method}

The power of light-emitting structure can be supplied by the driver chip or the power chip, as shown in Figure \ref{fig:2}. There has been some debate over whether AMOLED products use more power when powered by the driver chip or the power chip under light loading.

\begin{figure}[!t]
\centering
\includegraphics{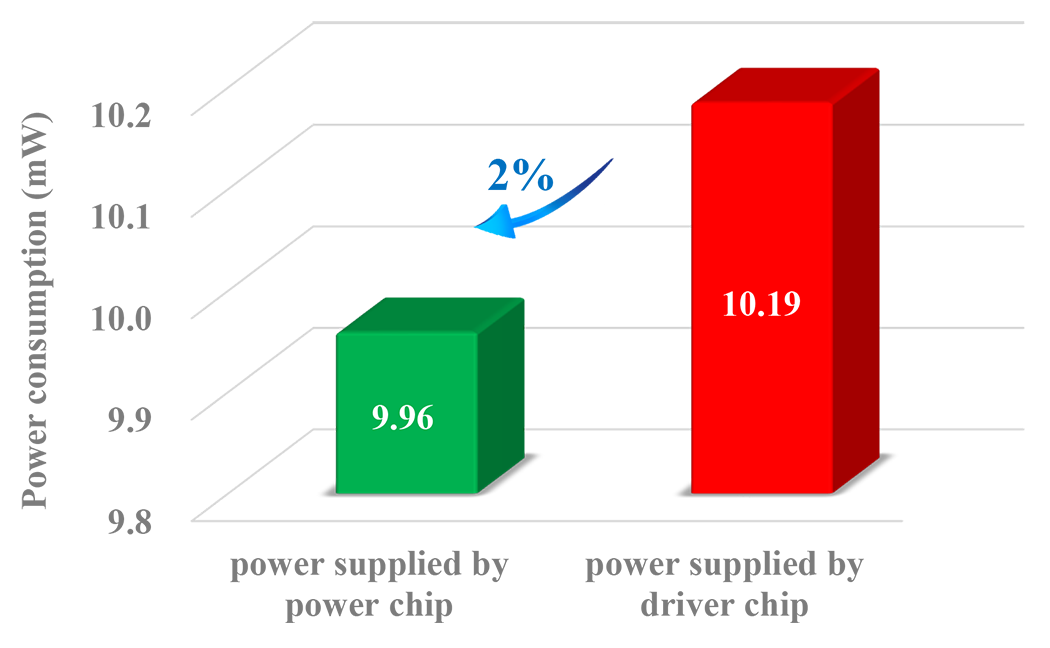}
\caption{The effect of power supplied by different kinds of chips on power consumption of one 1.43inch AMOLED display with the design of driver chip R and power chip A, under the condition of 10\% pixel on, 100nits and 15Hz.}
\label{fig:19}
\end{figure} 

\begin{figure}[!t]
\centering
\includegraphics{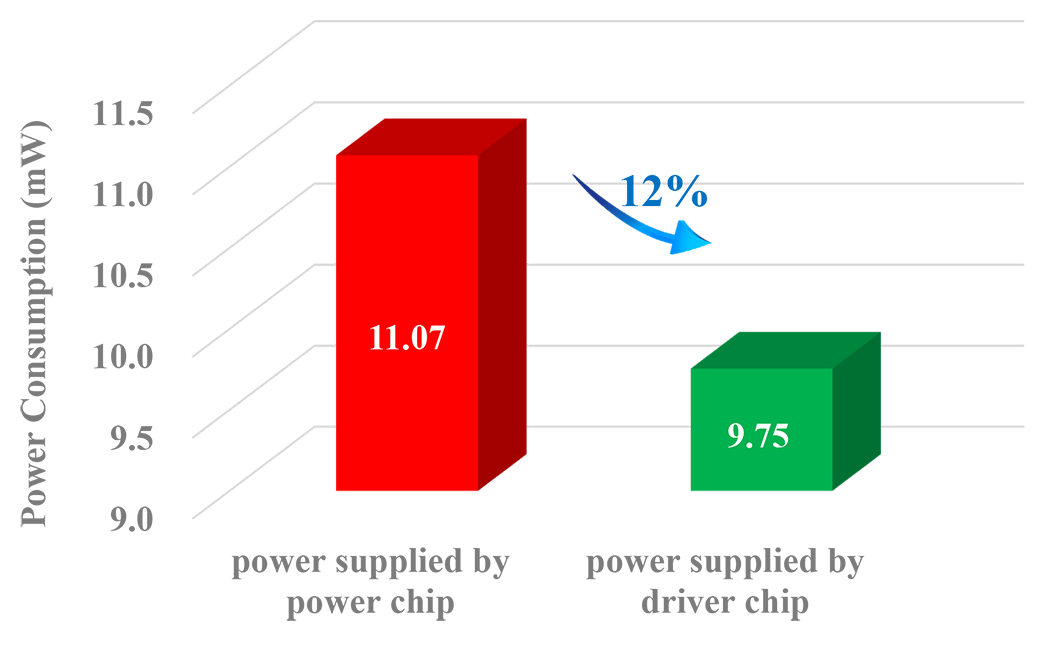}
\caption{The effect of power supplied by different kinds of chips on power consumption of one 1.43inch AMOLED display with the design of driver chip C and power chip C, under the condition of 10\% pixel on, 100nits and 15Hz.}
\label{fig:20}
\end{figure} 

This is related to the driver and power chips' light loading efficiency curves, as well as the definition of light loading. Figure \ref{fig:18} depicts the light loading efficiency curve, with an intersection point between the driver and power chip curves. And which power supply method is more power efficient is determined by the location of the light loading definition in relation to the efficiency intersection point.

Figure \ref{fig:18} shows that if the AMOLED display loading is in Case 1, it is more power efficient to be powered by the driver chip. In contrast, if the loading is in Case 2, it is more cost effective to use the power chip. The higher the loading, the greater the advantage of the power chip; conversely, the lower the loading, the greater the obvious advantage of the driver chip.

The selected chip type and combination influence the trend of the efficiency curve of light loading and the location of the intersection point. Different combinations of driver and power chips will produce varying results. There is no uniform conclusion on the optimal power supply method using different combinations of driver and power chip under the same operating conditions (loading) of the same product.

The power chip outperforms the driver chip in some cases (Figure \ref{fig:19}), while the driver chip outperforms the power chip in others (Figure \ref{fig:20}), according to the respective power conversion efficiency of the driver and power chips, as well as the display loading. As a result, the selection of power supply mode under light loading must be carefully considered in conjunction with the actual chip selection.

\begin{figure}[!t]
\centering
\includegraphics{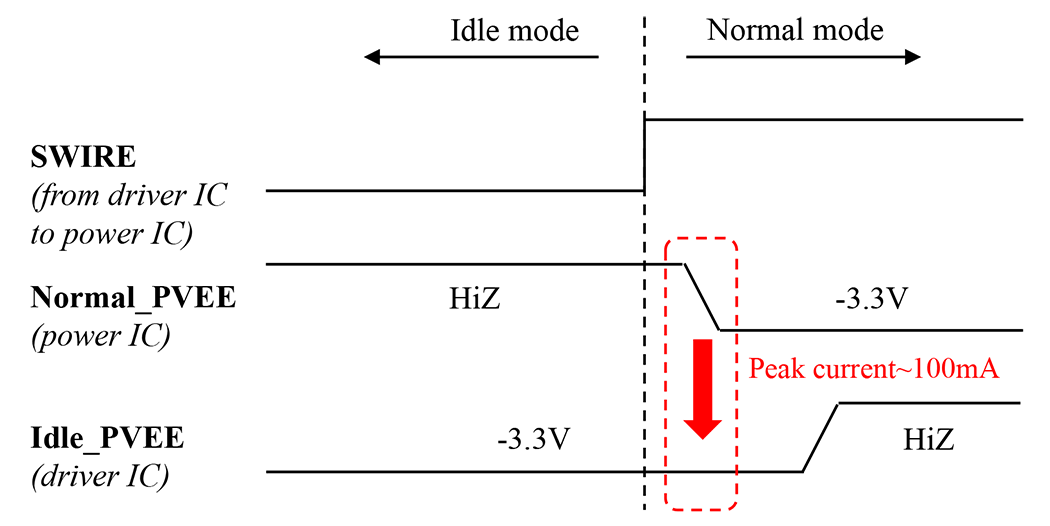}
\caption{Diagram of the risk of power on/off sequence between driver chip and power chip during mode switching from Idle to Normal mode.}
\label{fig:21}
\end{figure} 

\begin{figure}[!t]
\centering
\includegraphics[scale=0.9]{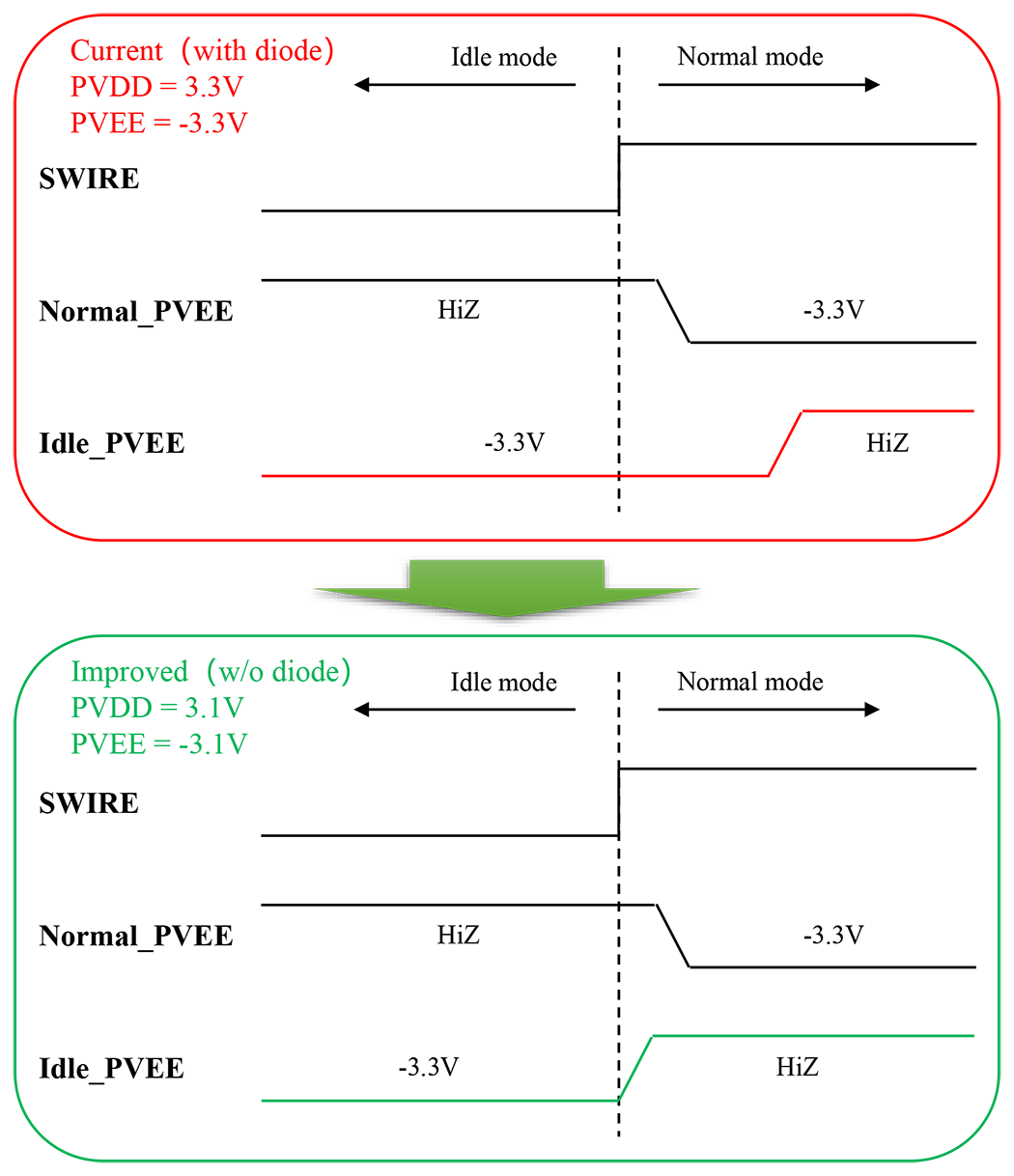}
\caption{Diagram of how to obtain a smaller operating voltage by eliminating the diode design by optimizing the power on/off sequence of driver chip.}
\label{fig:22}
\end{figure} 

\subsubsection{Intermodulation Mode Optimization}

When the power supply mode of an AMOLED display is switched between the driver chip and the power chip, the power on/off sequences of the chips must be matched. Because the driver chip and power chip on the market are designed by different companies, sequence matching may be risky (Figure \ref{fig:21}).

This risk will result in transient high current and instantaneous voltage jumps, which will not only affect the display effect when switching modes, but may also cause EOS damage to the chip. To avoid such issues, placing a diode between the power and driver chip can suppress the transient high current while maintaining normal display operation.

However, the characteristic curve of the diode device indicates that the diode will operate with voltage dividing phenomenon, implying that the voltage supplied has been partially divided by the diode. The diode design can be eliminated if the risk of transient high currents can be reduced by optimizing the power on/off sequence. Simultaneously, it enables the light-emitting structure to operate at a lower voltage, consuming less power.

As shown in Figure \ref{fig:22}, it turns off the idle PVEE slightly earlier than the Normal PVEE by optimizing the power on/off sequences of the driver and power chip, avoiding the period that may generate instantaneous high currents while having no significant visual impact.

\begin{figure}[!b]
\centering
\includegraphics{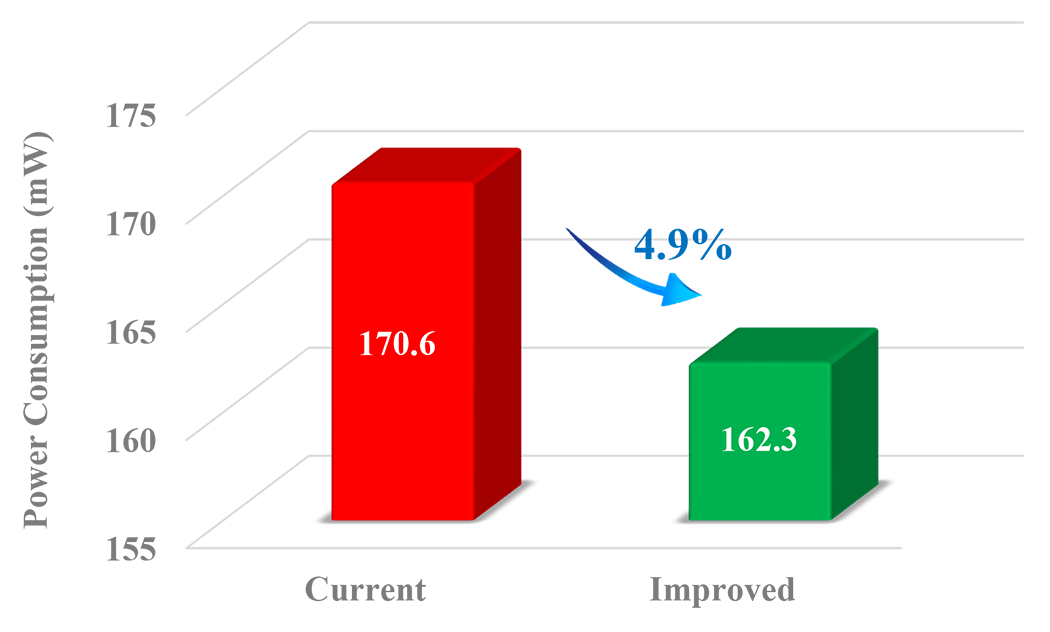}
\caption{The effect of design with and without diode by improving power on/off sequence of driver chip N on power consumption of one 1.75inch AMOLED display, under the condition of 100\% pixel on, 500nits and 60Hz.}
\label{fig:23}
\end{figure}

According to the measured results (Figure \ref{fig:23}), when the diode is removed, the trans-voltage of the light-emitting structure drops from 6.6V to 6.2V, resulting in a 5\% improvement in power consumption for the display module designed without PMIC. Because the conversion efficiency of PMIC differs slightly for different output voltages (see Chapter II), the improvement in power consumption for the display module designed with PMIC is slightly reduced.

\subsubsection{Automatic Optimization Of Configuration}

The power transmission path of the wearable product display is typically from the battery output VBAT to the power chip, then from the power chip output VCI to the driver chip, and finally from the driver chip to the display's various working voltages. We can follow the way in front of the power and driver chip voltage settings and voltage conversion method for reasonable optimization in the product design, so that the display is in the best efficiency point of the power supply.

However, there are many variables in the practical application. The loading of the display will change as the display content and operating mode change, the output voltage of the battery will drop as the power decreases, and other conditions that could not be considered at the beginning of the display design. Since the output voltage setting and generation mode of the driver and power chips are fixed, they cannot automatically respond to these changes and thus make the product deviate from the optimal efficiency point setting, resulting in unintended power loss in the display.

\begin{figure*}[!t]
\centering
\includegraphics[scale=0.9]{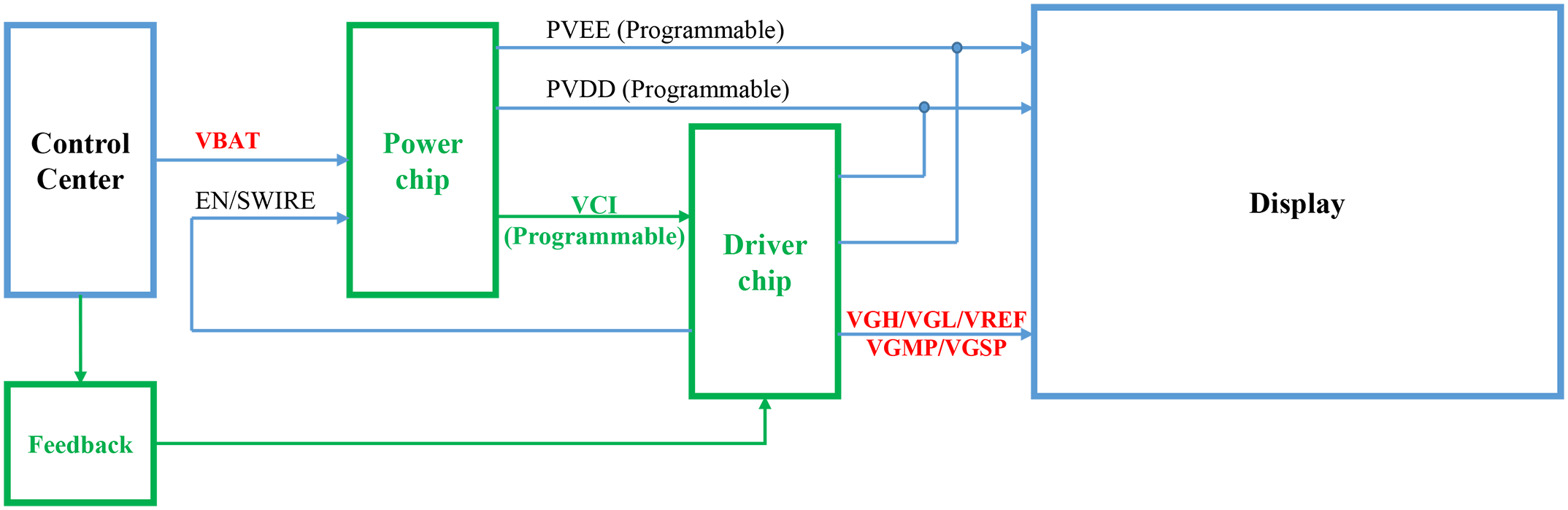}
\caption{Smart wearable AMOLED display power architecture with added feedback mechanism.}
\label{fig:24}
\end{figure*} 

As a result, a new power supply design scheme for AMOLED displays is suggested (Figure \ref{fig:24}). It automatically adjusts the operating state between the driver and the power chip based on external feedback. The driver chip adjusts all fixed output voltages to be programmed; at the same time, the driver chip accepts external information about display content, working status, battery level, and so on; and triggers the optimal output voltage and generation mode under different conditions based on the built-in instructions. It can ensure that the AMOLED display is always set to the lowest possible power consumption \cite{18}.

\section{Array Substrate}

\begin{figure}[!b]
\centering
\includegraphics[scale=0.9]{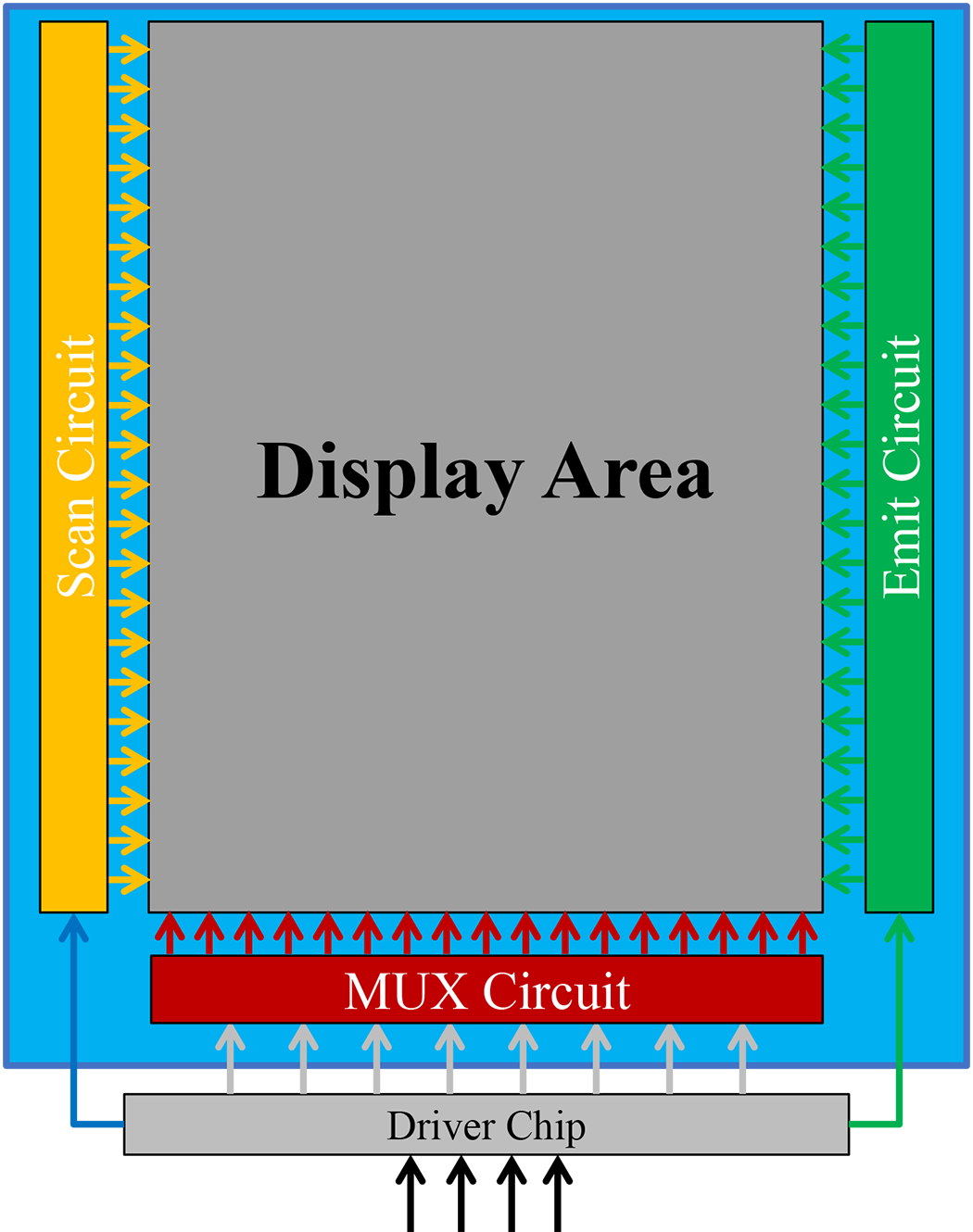}
\caption{Schematic diagram of array substrate structure}
\label{fig:25}
\end{figure}

The array substrate is the display's core component (Figure \ref{fig:25}), and the technical fields involved in its design and fabrication are all part of the field of microelectronics. A semiconductor thin film preparation process is used to fabricate the circuit structure on the substrate, resulting in a pixel matrix that can control the brightness level of each pixel autonomously. The substrate can be rigid glass or flexible plastic, depending on the application.

The array substrate's power consumption level is related to the process conditions used, the design size, and drive timing. Low temperature polysilicon (LTPS) is the current mainstream array substrate process for AMOLED products \cite{19,20,21,22}. The array substrate's preparation process essentially determines the approximate range of its power consumption level, and because the process equipment and process parameters are relatively fixed, it is difficult to achieve a more significant improvement in power consumption by process means without affecting product performance.

The power loss of the array substrate itself is a relatively small percentage of the overall product power consumption, resulting in a low cost-performance ratio of the overall AMOLED display power improvement by optimizing the array substrate's power consumption level. As a result, the relevant power improvement idea is modified to optimize the array substrate parameter settings in order to achieve the goal of reducing the power consumption of the driver chip.

Next, the logic of reducing the power consumption of the driver chip will be followed in order to further challenge the performance limits of the array substrate and try to balance the negative effects by adjusting the design and timing aspects.

\subsection{Voltage Setting Optimization}

According to the previous analysis of the power consumption improvement of the driver chip's analog part (see Chapter III-A), the driver chip's power consumption level can be effectively improved by optimizing the boost configuration and conversion mode. However, these optimizations are not chosen at random; the output specifications of the relevant voltages must meet the basic operating requirements of the array substrate, and a balance between power consumption levels and display effects must be established.

For example, we discovered a new set of voltage configurations that can keep the pixel circuit working properly based on the driver chip voltage generation settings and the pixel circuit voltage operation logic (Table \ref{tab:4} and Table \ref{tab:5}). When compared to the original voltage configuration, each voltage is adjusted in the manner in which the voltage is generated within the driver chip.

\begin{table}[!t]
\caption{Internal voltage configuration of driver chips R in different boost modes}\label{tab:4}
\centering
\begin{tabular}{ccc}
\toprule
Output& Conversion Setting& Voltage (V)\\
\midrule
VCI& input& 3.3\\
AVDD& Boost: $2 \times VCI \rightarrow LDO$& 6\\
VGMP& $AVDD \rightarrow LDO$& 5.2\\
VGSP& $AVDD \rightarrow LDO$& 0.8\\
PVDD& $VCI$& 3.3\\
PVEE& $-VCI$& -3.3\\
VCL& Boost: $-1 \times VCI$& -3.3\\
VREF& $VCL \rightarrow LDO$& -3.1\\
VGHR& Boost: $1 \times AVDD \rightarrow LDO$& 6\\
VGLR& Boost: $VCL-VCI \rightarrow LDO$& -6\\
\bottomrule
\end{tabular}
\end{table}

\begin{table}[!t]
\caption{The improved operating voltage setting based on the operating principle of the array substrate’s pixel circuit.}\label{tab:5}
\centering
\begin{tabular}{ccc}
\toprule
Output& Conversion Setting& Voltage (V)\\
\midrule
VCI& input& 3.3\\
AVDD& $VCI$& 3.3\\
VGMP& $AVDD \rightarrow LDO$& 3.2\\
VGSP& $AVDD \rightarrow LDO$& 0.8\\
PVDD& $VCI \rightarrow LDO$& 1.8\\
PVEE& $VCL$& -3.3\\
VCL& Boost: $-1 \times VCI$& -3.3\\
VREF& $VCL$& -3.3\\
VGHR& Boost: $2 \times AVDD \rightarrow LDO$& 6\\
VGLR& Boost: $VCL-VCI \rightarrow LDO$& -6\\
\bottomrule
\end{tabular}
\end{table}

\begin{figure}[!t]
\centering
\includegraphics{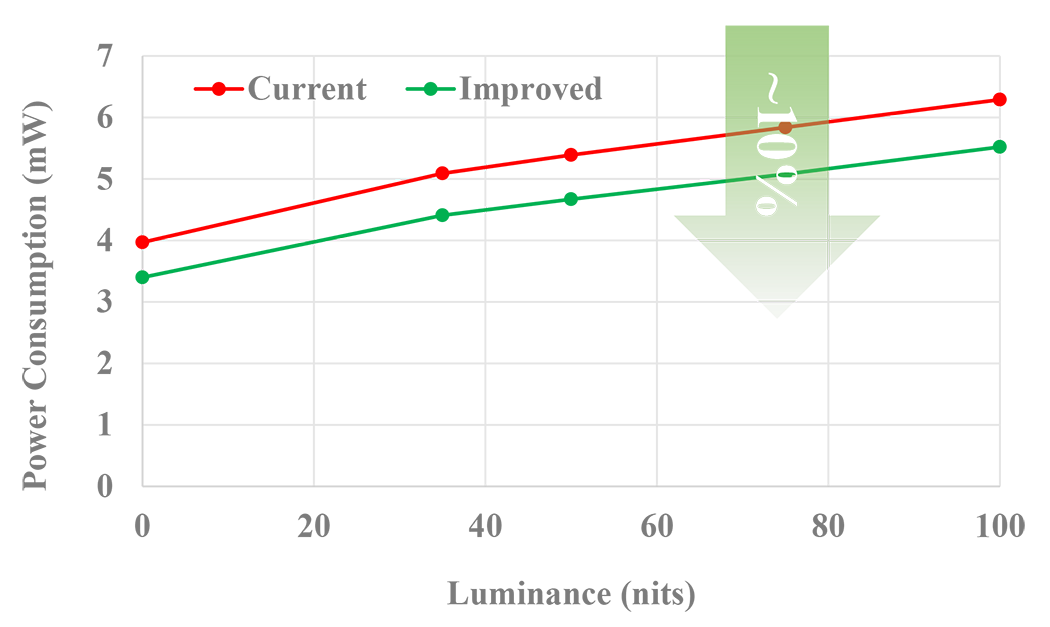}
\caption{The effect of driver chip N with different operating voltage settings on power consumption of one 1.41inch AMOLED display, under the condition of 10\% pixel on and 15Hz with different luminance.}
\label{fig:26}
\end{figure}

By sacrificing the power loss of the driver chip's secondary booster circuit (VGH) for the power savings of the primary booster circuit (AVDD), this improvement can eventually result in a 10\% power reduction (Figure \ref{fig:26}). However, because the pixel circuits used by each panel company differ slightly, they must be evaluated and verified in conjunction with their own actual conditions.

\subsection{Frequency Setting Optimization}

\begin{figure}[!b]
\centering
\includegraphics{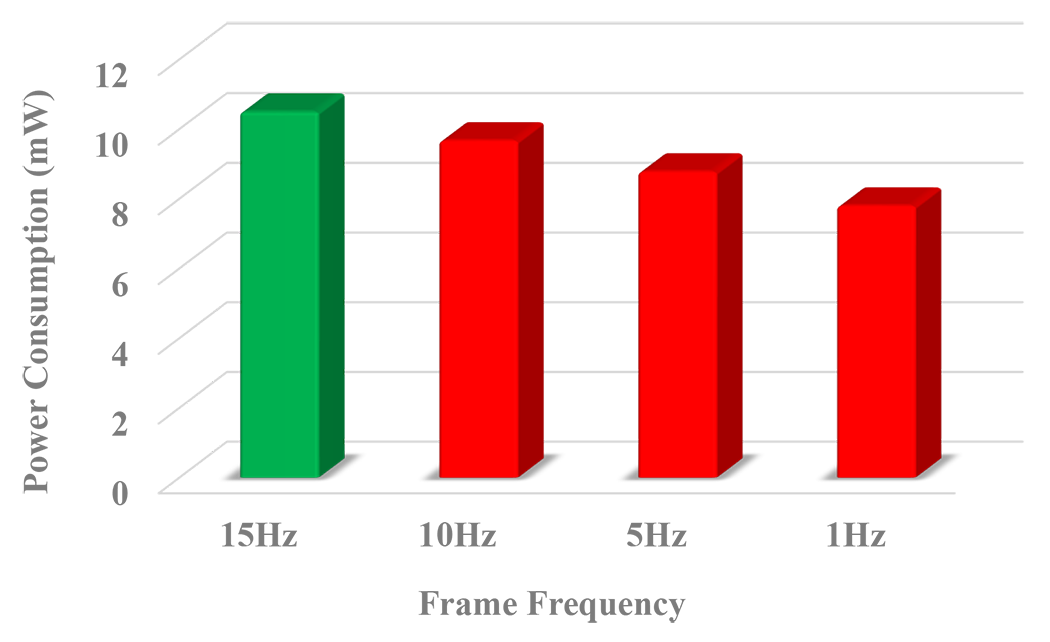}
\caption{The effect of driver chip R with different frame frequency setting on power consumption of one 1.32inch AMOLED display, under the condition of 10\% pixel on and 10nits.}
\label{fig:27}
\end{figure}

According to the measures to improve the power consumption of the driver chip's digital circuit (see Chapter III-B), lowering the refresh frequency is an immediate measure to improve power consumption. It is possible to design the driver chip to achieve a very low refresh frequency and lower power performance (Figure \ref{fig:27}).

However, due to the limitations of the array substrate, we are unable to do so \cite{23}. When the refresh frequency of the current mainstream LTPS process is reduced to less than 10Hz, flicker and retention such display problems appear, which are primarily caused by TFT characteristics and hysteresis effect. However, if the aforementioned issues can be resolved through design, 10Hz and lower refresh frequencies in the LTPS process will be possible, the power consumption of the driver chip will also be reduced.

\subsubsection{Flicker Improvement}

The flicker phenomenon occurs as a result of a change in display brightness that the human eye can detect as a result of light and dark \cite{24,25,26}. This change is primarily due to the fact that the brightness of the screen cannot be maintained to appear changed before the next display data refresh is completed. It is not easily recognized by the human eye at high frequencies, but becomes apparent when the frequency is reduced.

The leakage of TFT devices responsible for the switching function in the pixel circuit of the array substrate is the primary cause of the flicker phenomena, and this is the actual issue that the choice of the LTPS process had to deal with. By increasing the area of storage capacitors, decreasing the size of TFT devices responsible for driving function, and increasing the size of TFT devices responsible for switching function, it is possible to improve the TFT leakage phenomenon and improve the flicker phenomenon in the low frequency state of the display.

\subsubsection{Retention Improvement}

The retention phenomenon refers to the display when switching from dark to light, in which the brightness of the first frame after the switch does not reach the expected target and only gradually recovers later. The retention phenomenon is primarily related to the threshold voltage shift caused by the hysteresis effect of TFT device \cite{27,28}, which results in the brightness not reaching the expected target at the first time the screen is switched on.

Due to the long time interval between two frames of writing display data, the low refresh frequency screen is easier to recognize by human eyes. By quickly inserting a high-frequency screen when the low-frequency screen is switched, the time when abnormal brightness occurs can be compressed to the point where the human eye cannot recognize it, effectively improving the retention phenomenon at low frequencies.

\subsection{Process Improvement}

The current leakage characteristics of TFT devices are the main reason for the flicker phenomenon in low frequency state, and an important factor that prevents the display substrate of LTPS process to further reduce the refresh frequency. Through a series of design optimization can be compressed to less than 10Hz, but there is still a gap from the ultimate frequency that can be supported by the driver chip.

\begin{figure}[!t]
\centering
\includegraphics[scale=0.7]{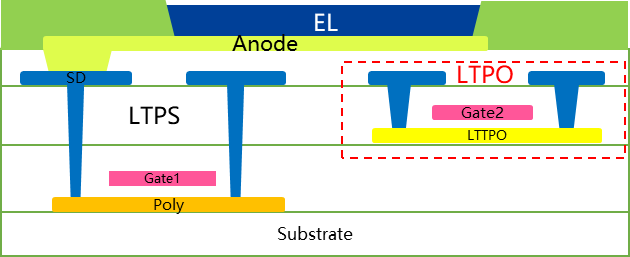}
\caption{LTPO device structure}
\label{fig:28}
\end{figure} 

\begin{figure*}[!b]
\centering
\includegraphics[scale=0.9]{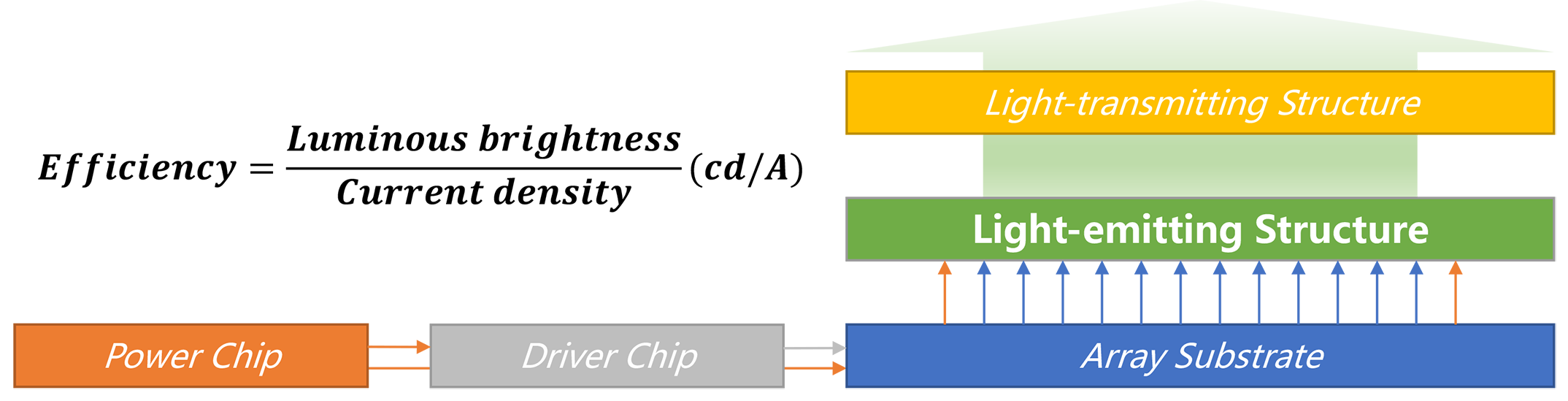}
\caption{Schematic diagram of light-emitting structure}
\label{fig:29}
\end{figure*} 

After all, the LTPS process still has limitations, and there are bottlenecks in the pursuit of lower refresh frequency on the road. The low temperature polycrystalline oxide (LTPO) process technology can be used to achieve a lower refresh frequency \cite{29,30,31}. The LTPS process is still used for the TFT devices responsible for driving function with high mobility requirements, while the LTPO process is used for the TFT devices responsible for switching function with low leakage requirements (Figure \ref{fig:28}).

At the same time, even if the TFT device leakage problem is solved by introducing the LTPO process at the ultra-low refresh frequency, the threshold voltage shift problem caused by the TFT hysteresis effect remains. The low frequency write and high frequency reset can effectively suppress the display abnormality caused by the threshold voltage drift, and the ultimate refresh frequency can be achieved under the LTPO process by adjusting the output timing of the driver chip \cite{32}.

\section{Light-emitting Structure}

The key signal conversion unit of AMOLED products is the light-emitting structure (Figure \ref{fig:29}), which completes the conversion of electrical signals to optical signals. The key parameter for measuring the light-emitting structure is conversion efficiency, and its size has a direct impact on the power consumption level.

\begin{figure}[!t]
\centering
\includegraphics[scale=0.7]{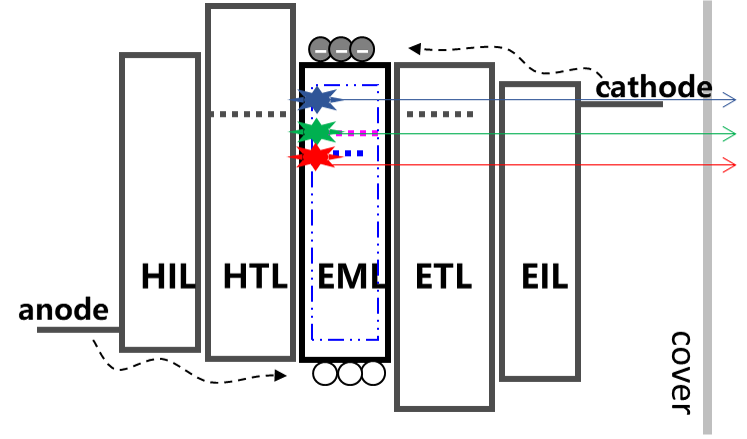}
\caption{OLED luminescence principle}
\label{fig:30}
\end{figure}

OLED is a kind of current-type organic light-emitting device, which is a phenomenon that causes luminescence through the injection and recombination of carriers, and the luminous intensity is proportional to the injected current (Figure \ref{fig:30}). Under the action of the electric field, the holes generated by the anode and the electrons generated by the cathode will move, and they will be injected into the hole transport layer and the electron transport layer respectively, and migrate to the light-emitting layer. When the two meet in the light-emitting layer, an energy exciton is generated, which excites the light-emitting molecule and eventually produces visible light.

OLED device efficiency is determined by their structure, material type, and process technology \cite{33,34,35}. It is also necessary to balance the relationship between efficiency, lifetime, and cross-voltage during the development process, which must be combined with specific application requirements to make trade-offs.

\subsection{Material Process Improvement}

The key factors that affect efficiency in OLED devices can be disassembled based on the light-emitting process. The relevant factors in the development of OLED devices can be adjusted to improve design efficiency. However, the vast majority of these key factors are important parameters determined during the development of the light-emitting structure's material system \cite{36,37} and cannot be easily changed. After determining the current power consumption target and projecting the efficiency gap, completely new material systems and device structures can be developed with this as the goal.

The development of material systems and device structures for light-emitting structures can only assess the level of efficiency improvement from a theoretical point of view, but the actual situation is related to the process technology of the light-emitting structures. The OLED device's manufacturing process has a significant impact on efficiency \cite{38,39,40}, which can be improved by focusing on optimizing the structure, reducing defects, removing impurities, flattening the interface, and preventing cracking.

\begin{figure}[!b]
\centering
\includegraphics[scale=0.7]{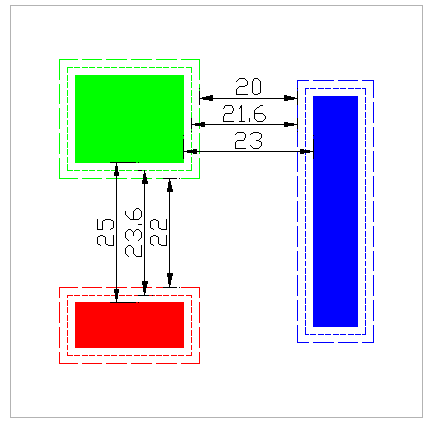}
\caption{The design of pixel definition layer}
\label{fig:31}
\end{figure} 

\subsection{Increase the Light Output Area}

Pixel definition layer (PDL) is one of the processes in the array substrate preparation process, and the design size of the light-emitting structure is directly determined by its design size (Figure \ref{fig:31}). According to the OLED device efficiency formula, if the PDL can be optimized to increase the light-emitting area, the luminous flux can be increased to increase the luminous brightness, and thus improve the efficiency in disguise. The lower the current density under the same optical target, the same light-emitting structure with a larger light-emitting area.

Optimizing the PDL size design not only to challenge the current Design Rule, but also to take into account the efficiency differences of the three different colors of OLED devices. To balance the efficiency difference and ensure the overall power consumption optimization level is maximized, the light-emitting area for the low efficiency colors should be expanded as much as possible, while the high efficiency colors can be appropriately reduced and avoided. However, due to design constraints and process bottlenecks, the aforementioned measures are currently less cost-effective improvements that can be used as a supplement.

\subsection{Dynamic Real-time Adjustment}

\begin{figure}[!t]
\centering
\includegraphics{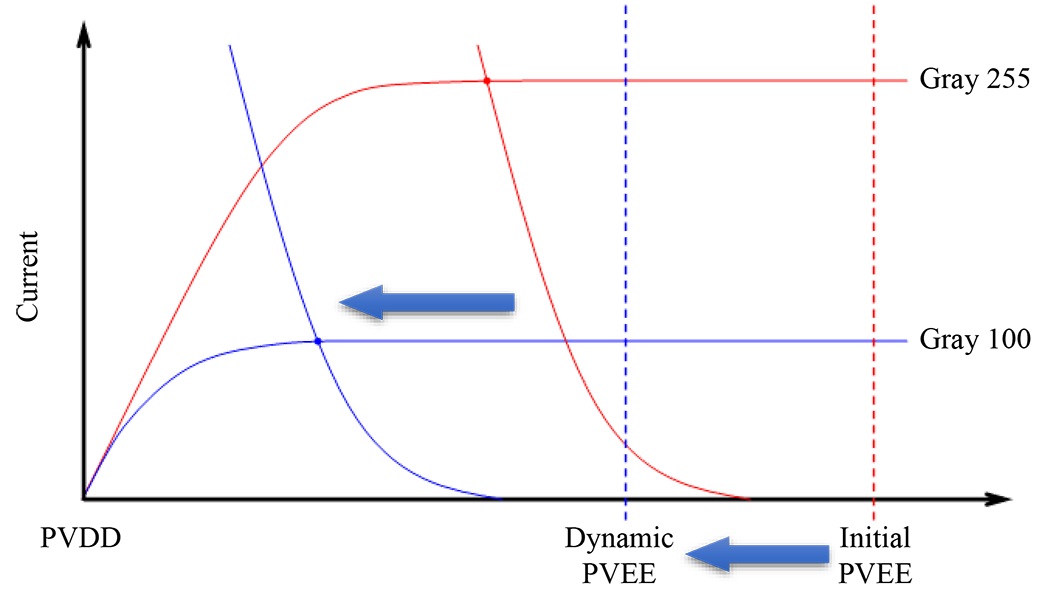}
\caption{The semiconductor characteristic curves of TFT driver tube and OLED diode}
\label{fig:32}
\end{figure}

According to the pixel circuit, the circuit structure between PVDD and PVEE voltage can be simplified to a series structure of TFT driver tube and OLED diode, and the intersection of the characteristic curves is shown in Figure \ref{fig:32}. The normal operation of AMOLED display requires the intersection point of the two characteristic curves to be located in the saturation area of the TFT characteristic curve. As the brightness of the display decreases, the current flowing through the OLED device decreases, and the saturation area of the TFT characteristic curve will have an inflection point moving forward. At this time, the voltage difference between PVDD and PVEE is properly reduced, which will not affect the optical output effect of the display.

From the standpoint of wearing the entire machine, because the content of the displayed screen is various, the high level of brightness is not always maintained. If the driver chip can recognize the maximum brightness of the display screen and control the power supply chip for dynamic adjustment based on the pre-programmed maximum brightness and adjustable PVEE voltage correspondence. At the same time, you can achieve accurate power consumption control based on the screen display content by using different output across the voltage power conversion efficiency of the more stable power supply chip \cite{41}.

\section{Light-transmitting Structure}

\begin{figure}[!t]
\centering
\includegraphics[scale=0.9]{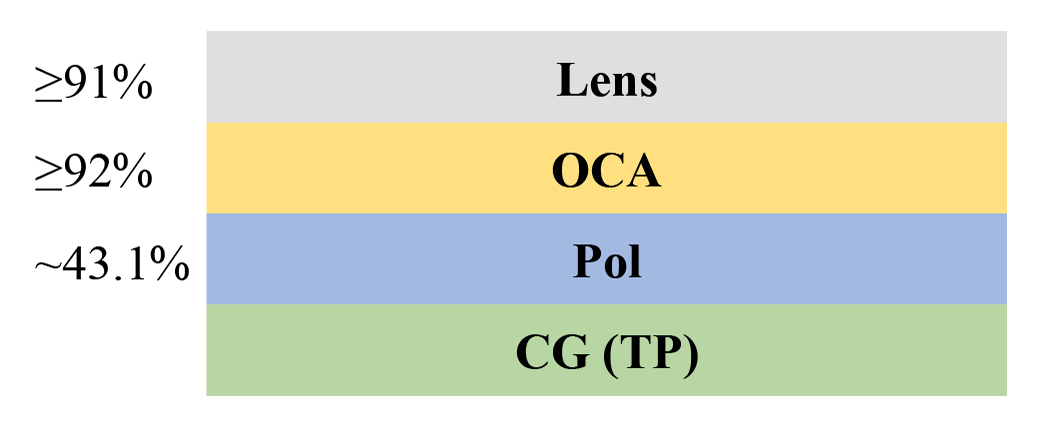}
\caption{Schematic diagram of light-transmitting structure}
\label{fig:33}
\end{figure}

The last key component of an AMOLED display is the light-transmitting structure, which is made up of optical components with various functions and roles. The transmission rate of the light-transmitting structure will determine how much of the final result can be presented after the previous series of signal transmission and conversion. Figure \ref{fig:33} depicts the light-transmitting structure of a conventional rigid glass substrate, which consists primarily of cover glass (CG), polarizer (Pol), optically clear adhesive (OCA), and lens glass.

Unfortunately, due to the current polarizer transmittance limitation, there is a more than 50\% loss of light as it passes through here. As a result, based on this traditional structure, the most frequently discussed topic when evaluating the reduction of product power consumption is how to improve the transmittance of polarizers.

A polarizer is an optical structure that alters the polarization of light \cite{42,43}. It is composed primarily of a linear polarizing film and a 1/4 phase delay film, and its primary function is to reduce the reflective effect of external incident light caused by reflection from the metal of array substrate. Based on this functional requirement, the related design and material selection result in a low polarizer transmittance.

Although it is foreseeable that improving polarizer transmittance can increase the brightness of the light output and thus improve efficiency in disguise to achieve a greater current reduction under the same product brightness, the resulting impact on reflectivity and cost increase has been a real problem that cannot be avoided.

The use of flexible substrates, as well as the maturity of thin-film encapsulation (TFE) technology, have opened up new avenues for improving light-transmitting structure transmittance. It has blurred the clear distinction between array substrate, light-emitting structure, and light-transmitting structure, broadening people's thinking and direction in addressing the problem.

\subsection{MLP Technology}

\begin{figure}[!t]
\centering
\includegraphics[scale=0.9]{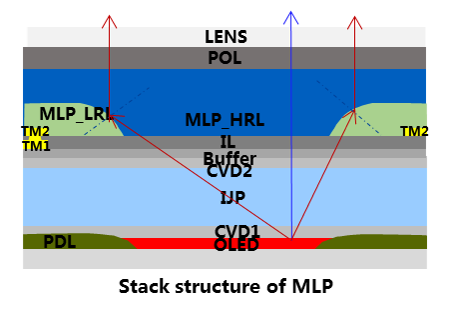}
\caption{Stack structure of MLP}
\label{fig:34}
\end{figure}

Micro lens panel (MLP) technology is a new attempt to improve the transmission rate of products based on the TFE packaging method \cite{44,45,46}. It uses low refractive materials to prepare micro lens patterns on each sub-pixel by photolithography, and introduces high-refractive materials in the planarization layer structure (Figure \ref{fig:34}).

It increases the efficiency of light output coupling through the lens interface between the low refractive and high refractive materials, so that the emitted light is refracted directly at the lens interface, resulting in light focusing.

MLP technology essentially sacrifices viewing angle for frontal brightness. Although power consumption will be significantly reduced, the indicator of view angle brightness decay will be negatively impacted.

\subsection{COE Technology}

\begin{figure}[!b]
\centering
\includegraphics{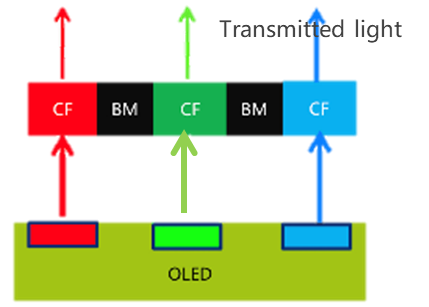}
\caption{Stack structure of COE}
\label{fig:35}
\end{figure}

Color filter on encapsulation (COE) technology is a polarizer replacement technology based on TFE technology \cite{47,48,49}. Instead of a polarizer, it uses a color filter structure prepared on the surface of the encapsulation to reduce the reflectivity of the display to external incident ambient light (Figure \ref{fig:35}). In the color filter, the black masking material intercepts the majority of the reflected light, while the black PDL material effectively masks the metal lines to reduce reflections.

Although COE technology does not yet have the same effect as polarizers in eliminating ambient light reflectivity, its contribution to power consumption reduction is significant. Simultaneously, when compared to the structures of conventional rigid glass substrate and MLP technology, COE technology can further reduce display thickness and is regarded as one of the key technologies for future flexible display products.

\section{Conclusion}

The power consumption level of wearable AMOLED displays is a systematic and comprehensive result, and the relevant influencing factors run through all aspects of the product from design to process. The relevant factors are previously interlocked with each other and need to be judged from a higher global perspective.

In the traditional electrical filed (power chip and driver chip), reducing additional losses, streamlining the conversion path, reducing the operating frequency is the main idea to reduce the power input; 

In the traditional optical filed (light-transmitting structure), streamlining the design of the optical path, selecting of high-transmission materials, reshaping the optical structure is the main idea to improve the optical output; 

In the electro-optical conversion filed (array substrate and light-emitting structure), reducing their own loading, optimizing the material matching, reducing process defects are the main ideas to improve efficiency.

The optimization of each field has to be carried out without affecting the overall performance of the product, and once there is a negative impact on other fields, a comprehensive solution needs to be considered.

From the perspective of smart wearable products as a whole, according to the application environment and scenario of the product, we can judge whether the specification requirements for the display are reasonable and necessary, and we can adjust the design scheme comprehensively according to the concession of the specification to achieve low-power design.

The power consumption optimization for each field is also not isolated, but can be complementary and linked to each other. Especially for the highly integrated driver chip, digital technology and artificial intelligence algorithms can be used to precisely control the working mode of the product according to the actual situation.


%





\ifCLASSOPTIONcaptionsoff
  \newpage
\fi



%

%

\begin{IEEEbiography}
[{\includegraphics[width=1in,height=1.25in,clip,keepaspectratio]{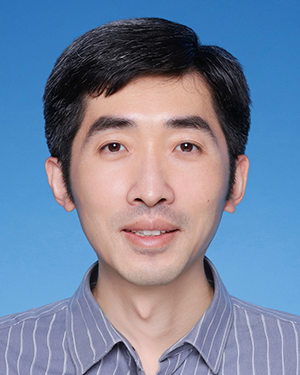}}]
{Bojia Lyu} was born in Shenyang, Liaoning, China, in 1982. He received B.S. degree in the major of electronic science and technology, and M.S. degree in the major of microelectronics and solid state electronics from Dalian University of Technology, Dalian, Liaoning, China, in 2005 and 2008, respectively. He is currently pursuing the Eng.D. degree in the major of electronic information with the School of Electronic Information and Electrical Engineering, Shanghai Jiao Tong University, Shanghai, China.

In 2008, he joined with Shanghai Tianma Microelectronics Co., Ltd., Shanghai, China. From 2010 to 2022, he worked as an Electronics Engineer of AMOLED design. During this period, he completed electronic, driving and algorithm development for AMOLED technology and led the electronic design of various AMOLED products for mobile phone and smart wearable application. He is currently a Professorate Senior Engineer with Shanghai Tianma Microelectronics Co., Ltd. His research interests include electronics design, driving system and image algorithm of display filed.

\end{IEEEbiography}







\end{document}